\documentclass{aa}
\usepackage{graphics}

\newcommand{\mi}{\mbox{$M_{\rm i}$}}
\newcommand{\diff}{\mbox{${\rm d}$}}

\begin{document}

\thesaurus{06(08.05.3; 08.16.4; 08.13.2; 08.03.1; 11.13.1) }

\title{The third dredge-up and 
	the carbon star luminosity functions in the 
	Magellanic Clouds} 

\author{Paola Marigo$^{1,2}$,  L\'eo Girardi$^{1}$, Alessandro Bressan$^{3}$}
\institute{
$^1$ Max-Planck-Institut f\"ur Astrophysik, Karl-Schwarzschild-Str.\
	1, D-85740 Garching bei M\"unchen, Germany \\
$^2$ Department of Astronomy, University of Padova,
	Vicolo dell'Osservatorio 5, I-35122 Padova, Italy \\
$^3$ Astronomical Observatory, Vicolo dell'Osservatorio 5, I-35122,
Padova, Italy         
}

\offprints{Paola Marigo \\ e-mail: paola@mpa-garching.mpg.de \\
                  \hspace{2truecm} marigo@pd.astro.it} 

\date{Received 20 October 1998 / Accepted 18 December 1998}

\maketitle
\markboth{Marigo, Girardi \& Bressan}{The third dredge-up and the 
C-star luminosity functions in the Magellanic Clouds}

\begin{abstract}

We investigate the formation of carbon stars as a function of the
stellar mass and parent metallicity.  Theoretical modelling is based
on an improved scheme for treating the third dredge-up in synthetic
calculations of thermally pulsing asymptotic giant branch (TP-AGB)
stars.  In this approach, the usual criterion (based on a constant
minimum core mass for the occurrence of dredge-up, $M_{\rm c}^{\rm
min}$) is replaced by one on the minimum temperature at the base of
the convective envelope, $T_{\rm b}^{\rm dred}$, at the stage of the
post-flash luminosity maximum.  Envelope integrations then allow 
determination of $M_{\rm c}^{\rm min}$ as a function of stellar mass,
metallicity, and pulse strength (see Wood 1981), thus inferring if
and when dredge-up first occurs. Moreover, the final possible shut
down of the process is predicted.

Extensive grids of TP-AGB models were computed by Marigo (1998ab)
using this scheme. In this paper, we present and discuss the
calibration of the two dredge-up parameters (i.e.\ efficiency
$\lambda$ and $T_{\rm b}^{\rm dred}$) aimed at reproducing the carbon
star luminosity function (CSLF) in the LMC, using TP-AGB models with
original metallicity $Z=0.008$.  In addition to this, the effects of
different input quantities on the theoretical CSLF are analysed.  It
turns out that the faint tail is almost insensitive to the history of
star formation rate (SFR) in the parent galaxy, in contrast to the
bright wing which may be more affected by the details of the recent
history.  Actually, we find that the faint end of the CSLF is
essentially determined by the temperature parameter $T_{\rm b}^{\rm
dred}$. Once the faint end is reproduced, the peak location is a
stringent calibrator of the efficiency parameter $\lambda$.  The best
fit to the observed CSLF in the LMC is obtained with $\lambda=0.50$,
$\log T_{\rm b}^{\rm dred} = 6.4$, and a constant SFR up to an age of
about $5 \times 10^{8}$ yr.  This recent drop of the SFR is invoked to
remove a slight excess of bright carbon stars otherwise predicted.

A good fit to the observed CSLF in the SMC is then easily derived from
the $Z=0.004$ models, with a single choice of parameters
$\lambda=0.65$, $\log T_{\rm b}^{\rm dred} = 6.4$, and a constant SFR
over the entire significant age interval. The result for $\lambda$ is
consistent with the theoretical expectation that the third dredge-up
is more efficient at lower metallicities.


\end{abstract}

%%%%%%%%%%%%%%%%%%%%%%%%%%%%%%%%%%%%%%%%%%%%%%%%%%%%%%%%%%%%%%%%%%%%%%%%%%

\section{Introduction}
\label{intro} 

During the TP-AGB phase, recurrent episodes of dredge-up can
significantly affect the surface chemical composition of low- and
intermediate-mass stars. These episodes are usually known as ``the
third dredge-up''. The basic mechanism of the mixing to the surface
can be explained on the basis of the standard structure equations
applied to stellar models in quasi-static evolution (Iben \& Renzini
1983). During the peak of every flash in the helium shell, part of the
nuclear energy is converted into thermal energy, and the remaining
spent as mechanical work for the expansion of the overlaying matter.
Material at the hydrogen-helium ($X-Y$) discontinuity and beyond is
pushed outward so that the H-burning shell is temporarily
extinguished. At the same time, the base of the convective envelope
moves inward because of the progressive local raising of the radiative
temperature gradient over the adiabatic one, triggered by the increase
of the energy flux and the concomitant decline of the temperature
consequent to the expansion.

Dredge-up occurs if the convective envelope can penetrate into the
inter-shell region containing material that has just undergone helium
burning during the flash.  A certain amount of inter-shell matter is
then brought up to the surface, and pollutes the envelope with newly
synthesised nuclear products.  The dredge-up material mainly consists
of helium, carbon, oxygen and traces of heavy elements produced by
s-process nucleosynthesis (see Lattanzio \& Boothroyd 1997 for an
extensive review of the topic).

The formation of single carbon (C-) stars is usually ascribed to the
enrichment of $^{12}$C surface abundance produced by the third
dredge-up. The topic is important for a series of reasons. First,
C-stars are present in many Local Group galaxies, indicating the
presence of relatively young stellar populations. The precise age and
metallicity intervals relevant for the formation of C-stars, however,
are far from being fully assessed from both theory and
observations. Secondly, these stars are crucial for the interpretation
of a series of chemical abundance data. For instance, it is believed
that most SiC grains found in primitive meteorites are formed in
extended carbon-rich envelopes of C-stars (Gallino et al.\ 1990). The
emission features detected in the infrared spectra from extended
circumstellar shells around AGB stars are significantly affected by
the composition of the dust grains (Ivezic \& Elitzur 1995).  The
variety of chemical abundance patterns in planetary nebulae probably
reflects the different efficiency and duration of dredge-up episodes
in the progenitor stars, as well as the possible burning of CNO
isotopes at the base of their envelopes (the so-called hot-bottom
burning or envelope burning; see Marigo et al.\ 1998 and references
therein).

In recent years many investigations have been devoted to the analysis
of the occurrence and related efficiency of the dredge-up, and to the
nucleosynthesis driven by thermal pulses (Boothroyd \& Sackmann 1988cd;
Frost \& Lattanzio 1996; Straniero et al.\ 1997; Lattanzio \&
Boothroyd 1997; Herwig et al.\ 1997; and references therein).  The
main challenge of the theory in this case is offered by the so-called
{\it carbon star mystery}, first pointed out by Iben (1981).
Observations of AGB stars in the Large and Small Magellanic Clouds
(LMC and SMC) indicate that the luminosity range in which carbon stars
are found extends down to rather faint magnitudes ($M_{\rm bol} \ga
-3$), thus suggesting the evolution from low-mass progenitors ($\mi
\sim 1 - 3 \; M_{\odot}$).  In contrast, complete AGB models
easily explain the formation of carbon stars at much higher
luminosities ($M_{\rm bol} < -5 - -6.5$), but fail to reproduce the
faint part of the observed distribution, since dredge-up of carbon is
hardly found to occur during the TP-AGB evolution of low-mass models.

Although some improvements have recently been attained (Frost \&
Lattanzio 1996; Straniero et al.\ 1997; Herwig et al.\ 1997) the
present understanding of the dredge-up process is still
unsatisfactory, mostly because of the uncertainties in the treatment
of the mixing process and the definition of convective boundaries.  In
Sect.~\ref{pardup} we give an overview of the present state of art
emerging from the available results of complete modelling of the third
dredge-up, and discuss the role of analytical TP-AGB calculations in
possibly giving useful theoretical indications. 

In the present study, {\it the treatment of the third dredge-up in
synthetic TP-AGB calculations has been significantly improved.  A
simple technique based on integrations of static envelope models is
used in order to infer if and when convective dredge-up is likely to
occur during the evolution of a TP-AGB star}.  In this way, we abandon
the assumption of a constant parameter $M_{\rm c}^{\rm min}$, usually
adopted in synthetic calculations (Groenewegen \& de Jong 1993; Marigo
et al.\ 1996a), and the dependence on the current stellar mass and
metallicity is taken into account.  The method is described in
Sect.~\ref{methdup}.

A primary target of theoretical models of the AGB phase is the
interpretation and reproduction of the observed CSLFs in the
Magellanic Clouds (Sect.~\ref{cslf}).  This is because a sole observable
holds a number of profound implications for different physical
processes occurring in AGB stars (e.g.\ nucleosynthesis, mixing and
surface chemical enrichment, mass-loss), and sets at the same time
severe constraints both on theoretical prediction of stellar
parameters (e.g.\ evolutionary rates, luminosities, lifetimes, initial
mass of carbon stars' progenitors), and possibly on properties of the
related stellar populations to which carbon stars belong (e.g.\ the
history of star formation). Section~\ref{theo_cslf} illustrates the
derivation of theoretical CSLFs.

In this study the formation of carbon stars is investigated with the
aid of the parameterized treatment of the third dredge-up. In
Sect.~\ref{calibr}, the dredge-up parameters (efficiency $\lambda$,
and base temperature $T_{\rm b}^{\rm dred}$ of the envelope at the
post-flash luminosity peak) are first determined by demanding that the
theoretical CSLF, calculated for initial metallicity $Z = 0.008$,
suitably fits the distribution observed in the LMC.  This calibration
is important not only for deriving theoretical indications on
the dredge-up process itself, but also because it gives more
reliability to the analysis on the surface chemical composition of AGB
stars and related applications (e.g.\ estimation of chemical yields,
and chemical abundances in planetary nebulae).  The sensitivity of
the results to the underlying history of star formation and the
initial mass function (IMF) is also explored.

The same procedure is then applied to find the proper pair of
parameters to reproduce the CSLF in the SMC (Sect.~\ref{smc}).  In
this case evolutionary calculations for initial metallicity $Z=0.004$
are used. The differences between the LMC and SMC distributions are
discussed. Section~\ref{conclu} summarises our main conclusions.

%%%%%%%%%%%%%%%%%%%%%%%%%%%%%%%%%%%%%%%%%%%%%%%%%%%%%%%%%%%%%%%%%
\section{The role of synthetic calculations}
\label{pardup}

%%%%%%%%%%%%%%%%%%%%%%%%%% TABLE %%%%%%%%%%%%%%%%%%%%%%%%%%%%
\begin{table*}
\centering
\caption{The third dredge-up in low-mass stars according to full
TP-AGB calculations.  For given stellar mass $\mi$, envelope abundances
(in mass fraction) of hydrogen $X$ and helium $Y$, and adopted
mixing-length parameter $\alpha$, the core mass
$M_{\rm c}^{\rm min}$ at the onset of dredge-up (if found) is given.
The value of the bolometric magnitude, $M_{\rm bol} (M \rightarrow
C)$, corresponds to the pre-flash maximum quiescent luminosity marking
the transition to carbon star. The estimates inside parenthesis refer
to the faintest luminosity during the post-flash luminosity dip.}
\label{litmcmin}
\begin{tabular}{cccccll}
\noalign{\smallskip}
\hline
\noalign{\smallskip}
\mi\ ($M_\odot$) & $Z$ & $Y$ & $\alpha$ & 
$M_{\rm c}^{\rm min}$ ($M_\odot$) &
$M_{\rm bol}(M \rightarrow C)$ & Source \\
\noalign{\smallskip}
\hline
\noalign{\smallskip}
 3.0  & 0.020  &  0.28 & 1.0 & no dredge-up & ---------      & 
Wood (1981) \\
 3.0  & 0.020  &  0.28 & 1.7 &  0.58        & --not given-- &   
Herwig et al.\ (1997) \\
 3.0  & 0.020  &  0.28 & 2.2 &  0.611       & $-5.66$       &  
Straniero et al.\ (1997) \\
 2.0  & 0.020  &  0.30 & 1.5 & no dredge-up & ---------     &
Gingold (1975) \\
 1.7  & 0.001  &  0.28 & 1.0 & no dredge-up & ---------        &   
Wood (1981) \\
 1.7  & 0.001  &  0.26 & 1.5 &  0.665       & $-5.11$~~$(-4.68)$ &  
Boothroyd \& Sackmann (1988d) \\
 1.5  & 0.001  &  0.28 & 1.0 & no dredge-up & ---------        &   
Wood (1981) \\
 1.5  & 0.001  &  0.20 & 1.5 & 0.605        & --------- $(-4.0)$ &
Lattanzio (1989) \\
 1.5  & 0.003  &  0.30 & 1.5 &  0.69        & --not given-- &   
Lattanzio (1987) \\
 1.5  & 0.003  &  0.20 & 1.5 &  0.62        & $-5.20$       &   
Lattanzio (1986)  \\
 1.5  & 0.004  &  0.25 & 1.6 & no dredge-up & ---------       &   
Vassiliadis \& Wood (1993) \\
 1.5  & 0.006  &  0.30 & 1.5 &  0.70        & --not given-- &   
Lattanzio (1986) \\
 1.5  & 0.020  &  0.30 & 1.0 & no dredge-up & ---------        &   
Lattanzio (1987) \\
 1.5  & 0.020  &  0.28 & 2.2 &  0.63        & $-5.65$       &   
Straniero et al.\ (1997) \\
 1.5  & 0.020  &  0.20 & 1.5 & 0.635        & --------- $(< -4.8)$ &
Lattanzio (1989) \\
 1.0  & 0.021  &  0.24 & 1.5 & no dredge-up & ---------     &
Sch\"onberner (1979) \\
 1.0  & 0.001  &  0.20 & 1.5 &  0.630       & --------- $(-4.3)$ &
Lattanzio (1989) \\
 0.8  & 0.001  &  0.26 & 3.0 &  0.566       & $-4.35$~~$(-3.59)$ &   
Boothroyd \& Sackmann (1988d) \\ 
 0.7  & 0.001  &  0.25 & 1.5 &  0.612       & --not given-- & 
Iben \& Renzini (1982) \\
 0.6  & 0.001  &  0.30 & 1.5 & no dredge-up & ---------      &
Gingold (1974) \\
\noalign{\smallskip}
\hline
\end{tabular}
\end{table*}
%%%%%%%%%%%%%%%%%%%%%%%%%%%%%%%%%%%%%%%%%%%%%%%%%%%%%%%%%%%%%%%%%

The usual analytical treatment of the third dredge-up requires the
knowledge of three basic input quantities:
%%%%%%%%%%%%%%%%%
\begin{itemize}
\item the minimum core mass for dredge-up to occur $M_{\rm c}^{\rm
min}$;
\item the dredge-up efficiency $\lambda$;
\item the chemical composition of the convective inter-shell.
\end{itemize}
%%%%%%%%%%%%%%%%%
A TP-AGB star can experience the third dredge-up when its core mass
has grown over a critical value $M_{\rm c}^{\rm min}$, as indicated by
theoretical analyses of thermal pulses (e.g.\ Wood 1981; Iben \&
Renzini 1983; Boothroyd \& Sackmann 1988cd; Lattanzio 1989). Estimates
of $M_{\rm c}^{\rm min}$ based on complete evolutionary calculations
are still a matter of debate, since this quantity depends on the
delicate interplay between model prescriptions (e.g.\ opacities,
nuclear reaction rates, convection theory, mass-loss, time and mass
resolution) and intrinsic properties of the star (e.g.\ chemical
composition, envelope mass, strength of the thermal pulses).  In
Table~\ref{litmcmin} we report some values of $M_{\rm c}^{\rm min}$
taken from different authors. It turns out that the results vary
remarkably, even among models with the same total mass and
metallicity. The third dredge-up is often not found in low-mass
models, and if it is, the corresponding metallicities are quite low
and the transition luminosities 
[see the $M_{\rm bol}(M \rightarrow C)$ entry]
are in most cases still brighter than the faint wing of the CSLFs in
the Magellanic Clouds.  Nevertheless, a certain trend can be extracted
from the comparison of the results, i.e.\ the onset of the third
dredge-up is favoured at higher stellar mass $M$, lower metallicity
$Z$, and greater efficiency of envelope convection (obtained e.g.\ by
increasing the mixing-length parameter $\alpha$).

The parameter $\lambda$ expresses the dredge-up efficiency. It is
defined as the fraction of the core mass increment over the preceding
inter-pulse period, $\Delta M_{\rm c}$, which is dredged-up to the
surface by the downward penetration of the convective envelope:
\begin{equation}
\lambda = \frac{\Delta M_{\rm dredge}}{\Delta M_{\rm c}}  
\label{lambda}
\end{equation}
where $\Delta M_{\rm dredge}$ is the mass of the dredged-up material.

The question of the true value of $\lambda$ is also troublesome. The
effective penetration of the envelope is crucially affected by the
treatment of the instability of matter against convection, still
poorly understood.  In the standard approach, the boundary of a
convective region is defined by the Schwarzschild criterion, i.e.\
where the inward gravity force is balanced by the buoyancy force.
However, at this point the velocity of the convective eddies is not
null, so that they can effectively penetrate into formally stable
regions. This phenomenon is usually known as convective overshoot.  Over
the years, several attempts have been made in order to include this
effect into evolutionary calculations (see Chiosi et al.\ 1992 and
references therein).  However, at present it appears that there is no
real agreement among different authors on the criterion to define the
maximum inward penetration of the envelope at thermal pulses during
the AGB. Moreover, the results seem to depend quite sensitively on technical
and numerical details (see Frost \& Lattanzio 1996 for a discussion of
this aspect).

The heterogeneous situation is well represented by the two following
examples. On the one side, Straniero et al.\ (1997) obtain rather
efficient dredge-up just using the Schwarzschild criterion and without
invoking any extra-mixing, but drastically increasing both time and
mass resolution during calculations. On the other hand, Herwig et al.\
(1997) find deep extra-mixing during dredge-up on the basis of a
parameterized description of the overshoot region, built up to mimic
the indications from hydrodynamical simulations (Freytag et al.\ 1996)
of convection in stars of a different kind (i.e.\ main sequence stars
and white dwarfs).

As far as the efficiency of the third dredge-up is concerned, it is
not easy to summarise the results from different authors.  However,
suffice it to recall that recurrent estimates, derived from TP-AGB
calculations of low-mass stars, give at most $\lambda \sim 0.3 -
0.4$ and often even smaller (Iben \& Renzini 1984; Lattanzio 1986,
1987, 1989; Boothroyd \& Sackmann 1988d; Straniero et al.\ 1997).

In conclusion, the present state of art is the following.  On the one
hand, most authors still do not find significant dredge-up in their
calculations (Vassiliadis \& Wood 1993; Wagenhuber 1996; see also
Table~\ref{litmcmin}), or assume the efficiency $\lambda$ as an input
quantity (Forestini \& Charbonnel 1997).  On the other hand, the
difficulty of a too poor dredge-up in low-mass stars seems to be
overcome to some extent (Boothroyd \& Sackmann 1988d; Frost \&
Lattanzio 1996; Straniero et al.\ 1997; Herwig et al.\ 1997), but
extensive calculations over a wide range of stellar masses and
metallicities are required to test the results by means of a
meaningful comparison with observations.  Moreover, results are quite
different even for similar initial conditions (e.g.\ $M$ and $Z$).

A powerful investigation tool complementary to full calculations is
offered by synthetic TP-AGB models, which have been notably improved
and up-graded in recent years (Groenewegen \& de Jong 1993; Marigo et
al.\ 1996a, 1998; Marigo 1998ab).  As far as the third dredge-up is
concerned, synthetic TP-AGB calculations clearly suggest that a higher
efficiency is required (e.g.\ $\lambda \sim 0.65 - 0.75$) in order
to reproduce observational constraints (e.g.\ the CSLF in the LMC), a
goal still missed by complete modelling of the TP-AGB phase.

The use of the parameters $M_{\rm c}^{\rm min}$ and $\lambda$ plays a
relevant role in analytical models of the TP-AGB phase.  They both
significantly affect the predicted luminosity functions of carbon
stars, in particular the position of the peak and the low-luminosity
tail of the distribution.  Lowering $M_{\rm c}^{\rm min}$ favours the
formation of carbon stars of lower masses and at the same time results
in a longer carbon star phase; increasing $\lambda$ corresponds to a
higher fraction of low mass stars which become carbon stars at fainter
luminosities.  In Marigo et al.\ (1996a) it was simply assumed that
both $M_{\rm c}^{\rm min}$ and $\lambda$ were constant, thus
neglecting their likely dependence on the stellar metallicity, total
mass, core mass, and mixing length parameter.  Though being a rough
approximation, such a choice was meant to provide useful theoretical
indications.  The calibration of the dredge-up parameters was based on
the goal of reproducing the observed CSLFs in the LMC.  It resulted
that the suitable values were $M_{\rm c}^{\rm min} = 0.58\; M_{\odot}$
and $\lambda = 0.65$. These values are very close to those derived by
Groenewegen \& de Jong (1993). Basing on these results, Forestini \&
Charbonnel (1997) fixed $\lambda = 0.6$ in their full calculations, in
order to estimate the chemical yields from intermediate-mass stars.

%%%%%%%%%%%%%%%%%%%%%%%%%%%%%%%%%%%%%%%%%%%%%%%%%%%%%%%%%%%%%%%%%
\section{The minimum core mass for dredge-up}
\label{methdup}

In this section we illustrate the adopted scheme which replaces the
assumption of a constant $M_{\rm c}^{\rm min}$. To start with, it is
necessary to consider some indications about the dredge-up provided by
detailed calculations of thermal pulses.  They show that the maximum
inward penetration of the envelope convection, over the pulse cycle,
occurs when the surface luminosity attains its characteristic
post-flash peak, $L_{\rm P}$ (Wood 1981; Wood \& Zarro 1981; Boothroyd
\& Sackmann 1988d).  Dredge-up takes place if the envelope convection
zone is able to penetrate into the carbon enriched region previously
occupied by the inter-shell convection zone during the flash.

In Sects.~\ref{lpeak} and \ref{tdred} below we recall and discuss the
features which are relevant to our analysis on the occurrence of the
dredge-up, namely the behaviour of the peak luminosity $L_{\rm P}$,
and the temperature at the base of the convective envelope during the
dredge-up events.  In Sect.~\ref{mcmin} the method to infer $M_{\rm
c}^{\rm min}$ for each AGB model is described, and the results for
several values of the stellar mass and metallicity are presented.
Section~\ref{chemdred} details the adopted inter-shell composition.

\subsection{The $M_{\rm c}-L_{\rm P}$ relation}
\label{lpeak}

Wood \& Zarro (1981) and 
Wood (1981) pointed out that at the surface luminosity maximum the
outward-flowing luminosity, $L_r$, is almost constant (to a fraction
of a percent) throughout the envelope, from the top of the remnant
inter-shell convection zone to the surface. This is because 
the thermal timescale in the envelope (i.e.\ for
absorption/release of internal-gravitational energy during the
hydrostatic re-adjustment following the flash) is much shorter 
(i.e. few years) than
the evolutionary timescale for the envelope structure to be
changed (i.e. decades to centuries). 
The constancy of $L_r$ may no longer be true when significant
energy absorption can occur in the envelope, e.g.\ in stars with quite
massive envelopes ($M_{\rm env} > 2.5 \; M_{\odot}$), or large core
masses ($M_{\rm c} > 0.9 \; M_{\odot}$) (Wood \& Zarro 1981).

The fact that at the post-flash luminosity maximum the envelope is
close to hydrostatic and thermal equilibrium (Wood 1981), together with
the existence of a radiative buffer over the burning shell, constitute
the necessary physical conditions which secure the validity of the
so-called  zero-order-radiative-approximation (Schwarzschild
1958; Eggleton 1967; Pac\-zy\'ns\-ki 1970; Wood \& Zarro 1981).  In
brief, theory shows that $L_{\rm P}$ is essentially controlled by the
core mass of the star independently of
the outer envelope mass, in analogy with the existence of the $M_{\rm
c}-L$ relation during the quiescent inter-pulse periods for low-mass
AGB stars.  For more details on the basic conditions that
underlie the theory of a core mass-luminosity relationship, the
reader is referred to Refsdal \& Weigert (1970), Kippenhahn (1981)
and Tuchman et al.\ (1983); see also Boothroyd \& Sackmann (1988b) 
and references therein.

In this study we adopt the $M_{\rm c} - L_{\rm P}$ relation presented
by Wagenhuber \& Groenewegen (1998), derived from extensive
calculations of complete AGB models (Wagenhuber 1996). It expresses
the post-flash luminosity maximum as the sum of two terms, i.e.\
$L_{\rm P} = L_{(\rm P,full)} + \Delta L_{(\rm P,first)}$, where
\begin{eqnarray}
\label{lp}
L_{(\rm P,full)} & = & \left[ 93000 + 2758 \log(Z/0.02) \right]
\times \\\nonumber
& & \left\{ M_{\rm c}-[0.503 + 0.82 (Z-0.02)] \right\} 
\end{eqnarray}
and
\begin{equation}
\label{dlp} 
\Delta L_{(\rm P,first)} = 
	-10^{ [ 1.77+2.76 M_{\rm c,0}-(34.0+32.4 M_{\rm c,0})
\Delta M_{\rm c}]} 
\end{equation}
The former term [Eq.~(\ref{lp})] gives the post-flash luminosity maximum
(in $L_{\odot}$) for full-amplitude thermal pulses as a function of
$M_{\rm c}$ (in $M_{\odot}$), with a dependence on the metallicity $Z$
(in mass fraction).  The latter term [Eq.~(\ref{dlp})] represents a
negative correction which has to be added to $L_{(\rm P,full)}$ in
order to reproduce the fainter maximum in the light-curve (at given
$M_{\rm c}$) for the first pulses, expressed as a function of the core
mass at the first He-shell flash, $M_{\rm c,0}$, and the current core
mass increment, $\Delta M_{\rm c} = M_{\rm c}-M_{\rm c,0}$.

\subsection{A characteristic temperature for dredge-up}
\label{tdred}

Following Wood (1981), at the post-flash luminosity peak the envelope
convection reaches its maximum inward penetration (in mass fraction)
and base temperature, $T_{\rm b}^{\rm max}$. At the same time, the
nuclearly processed material involved in the He-shell flash is pushed
out and cooled down to its minimum temperature over the flash-cycle,
$T_{\rm N}^{\rm min}$.

To this respect, a very interesting and useful indication is given by
Wood (1981), who reported that at the stage of post-flash luminosity
maximum, the temperature $T_{\rm N}^{\rm min}$ always approaches a
limiting constant value ($\log T_{\rm N}^{\rm min} = 6.7 \pm 0.1$),
for all core masses studied ($0.57 \; M_{\odot} \le M_{\rm c} \le 0.88
\; M_{\odot}$), regardless of the metallicity ($Z=0.001$, $Z=0.01$,
and $Z=0.02$).

A similar remark can be found in the work by Booth\-royd \& Sackmann
(1988d) about the production of low-mass carbon stars. In the final
discussion the authors stated that ``it is probably no coincidence''
that in all cases considered (i.e.\ full AGB calculations for stars
with total mass and metallicity [$\mi=0.8 \; M_{\odot}, Z=0.001$],
[$\mi=1.2 \; M_{\odot}, Z=0.0022$], [$\mi=1.7 \; M_{\odot}, Z=0.001$],
and [$\mi=2.0 \; M_{\odot}, Z=0.001$]) the first episode of convective
dredge-up is characterised by a temperature at the base of the
convective envelope of approximately the same value, $\log T_{\rm
b}^{\rm max} \sim 6.5$. In other words, $T_{\rm b}^{\rm max} \sim
T_{\rm N}^{\rm min}$ at the first occurrence of the third dredge-up.

In conclusion, detailed analyses of thermal pulses suggest that the
third dredge-up takes place when the base temperature attains or
exceeds some critical value. Hereinafter, we will refer to it as 
$T_{\rm b}^{\rm
dred}$. Therefore, we adopt the following
criterion:
\begin{itemize}
\item
if $T_{\rm b}^{\max} < {T}_{\rm b}^{\rm dred} \Rightarrow$
no dredge-up can occur 
\item 
if $T_{\rm b}^{\max} \ge {T}_{\rm b}^{\rm dred} \Rightarrow$
dredge-up occurs
\end{itemize}
On the basis of the above considerations, it is natural to assume that
the fulfilment of the condition
\begin{equation}
\label{cond_mcmin}
T_{\rm b}^{\rm max} = {T}_{\rm b}^{\rm dred} 
\end{equation}
for the first time during the evolution marks the onset of dredge-up.

\subsection{ $M_{\rm c}^{\rm min}$ from envelope integrations}
\label{mcmin}

Thanks to the fact that at the time of the maximum penetration of the
external convection the envelope is in hydrostatic and thermal
equilibrium (Sect.~\ref{lpeak}), and that $T_{\rm b}^{\rm max} \ge
{T}_{\rm b}^{\rm dred}$ when dredge-up occurs (Sect.~\ref{tdred}), it
follows that the question of the minimum core mass for dredge-up can
be analysed with the aid of a stationary envelope model.

Envelope integrations are performed on the basis of the following
scheme.  According to the analytical demonstration of Tuchman et al.\
(1983) it turns out that in the zero-order-radiative-approximation
(see also Schwarzschild 1958; Eggleton 1967; Paczy\'nski 1970) 
the product of the temperature
$T_{\rm c}$ and radius $R_{\rm c}$ just above the burning shell,
depends on the surface luminosity, core mass $M_{\rm c}$, and chemical
composition of the radiative transition region:
\begin{equation}
\label{t7r}
\left(\frac{T_{\rm c}}{10^7}\right) R_{\rm c} = 
	\frac{2.291}{(5X + 3 - Z)} \left[ M_{\rm c} - 0.156 (1 + X)
	\frac{L}{10^4} \right]
\end{equation}
where $T_{\rm c}$ is given in K, $L_{\rm c}$, $M_{\rm c}$ and $R_{\rm
c}$ are expressed in solar units, $X$ and $Z$ are the abundances (in
mass fraction) of hydrogen and metals, respectively.  It is worth
specifying that the term $(5X + 3 - Z)$ comes from the definition of
the mean molecular weight under the assumption of a fully ionised gas,
$\mu = 4/(5X + 3 - Z)$, and the term $(1 + X)$ derives from the
expression of the opacity given by the Thomson scattering of
electrons, $\kappa = 0.2(1+X)$.

Equation~(\ref{t7r}) offers the boundary condition adopted here to
determine the envelope structure of an AGB star at the
luminosity-peak, i.e.\ the radius $R_{\rm c}$ of the spherical layer
below which the mass $M_{\rm c}$ is contained, must coincide with:
\begin{equation}
\label{rcor}
R_{\rm c} = \left(\frac{10^7}{T_{\rm c}}\right) 
	\frac{2.291}{(3 - Z_{\rm sh})} 
	\left[M_{\rm c} - 0.156 \frac{L_{\rm P}}{10^4}\right]
\end{equation}
Equation~(\ref{rcor}) is derived from Eq.~(\ref{t7r}) with $L=L_{\rm
P}$ as given by Eqs.~(\ref{lp}) and (\ref{dlp}), setting $X=0$ since
in this particular case we deal with material that has already
undergone complete H-burning. The metallicity $Z_{\rm sh}$ is that of
the inter-shell, i.e.\ $Z_{\rm sh} = X(^{12}{\rm C}) + X(^{16}{\rm O})
+ Z \sim 0.24$ according to the prescription given in
Sect.~\ref{chemdred}.  Moreover, we assume a constant value of the
temperature, $T_{\rm c}=10^8$~K, which gives a reasonable order of
magnitude of the minimum temperature for the ignition of helium.  The
sensitivity of the results to $T_{\rm c}$ is discussed in
Sect.~\ref{consistency} below.

Then, for given stellar mass, core mass, surface chemical composition,
and peak-luminosity $L_{\rm P}$, envelope integrations are performed
until we find the value of the effective temperature, $T_{\rm eff}$,
such that $M(R_{\rm c}) = M_{\rm c}$.  At this point, the structure of
the envelope is entirely and uniquely determined.

%%%%%%%%%%%%%%% TABLE %%%%%%%%%%%%%%%%%%%%%%%%%%
\begin{table*}
\caption{Results from envelope integrations for a ($2.0 \; M_{\odot},
Z=0.004$) model at the stage of the post-flash luminosity maximum for
given ($L_{\rm P}, M_{\rm c}$) pair.  Relevant quantities are shown as
a function of different choices of the input parameter $T_{\rm c}$ [see
Eq.~(\ref{t7r})].  The minimum core mass for dredge-up, $M_{\rm c}^{\rm
min}$, is derived assuming $\log T_{\rm b}^{\rm dred} = 6.4$.}
\label{hbbtab}
\centering
\begin{tabular}{ccccccc}
\hline
\noalign{\smallskip}
$M_{\rm c}/M_{\odot}$ & $\log(L_{\rm P}/L_{\odot})$ & $\log(T_{\rm
c}/10^7\,{\rm K})$ & $\log(R_{\rm c}/{\rm cm})$ & 
$\log(T_{\rm eff}/{\rm K})$ & 
$\log(T_{\rm b}^{\rm max}/{\rm K})$ & $M_{\rm c}^{\rm
min}/M_{\odot}$ \\
\noalign{\smallskip}
\hline
\noalign{\smallskip}
 0.5400 & 3.6598 & 0.8 & 9.6328 & 3.580637 & 6.48485 & 0.52551 \\
        &        & 1.0 & 9.4328 & 3.580642 & 6.48461 & 0.52556 \\
        &        & 1.2 & 9.2328 & 3.580648 & 6.48440 & 0.52563 \\
        &        & 1.4 & 9.0328 & 3.580654 & 6.48408 & 0.52569 \\
        &        & 1.6 & 8.8328 & 3.580660 & 6.48379 & 0.52576 \\
        &        & 1.8 & 8.6328 & 3.580667 & 6.48346 & 0.52584 \\
\noalign{\smallskip}
\hline
\end{tabular}
\end{table*}
%%%%%%%%%%%%%%%% TABLE %%%%%%%%%%%%%%%%%%%%%%%%%%

\subsubsection{Consistency checks}
\label{consistency}

The consistency of our prescriptions must be checked.  First, the
assumption of a constant $T_{\rm c}=10^8$~K seems to be arbitrary.
However, we checked that varying this temperature by as much as 1
dex -- far more than allowed by reasonable changes in the He-burning
reaction rates -- produces quite negligible changes on the results of
envelope integrations (e.g.\ the values of $T_{\rm b}^{\rm max}$,
$T_{\rm eff}$, $M_{\rm c}^{\rm min}$; see Table~\ref{litmcmin}).  Such
invariance is just expected, given the physical decoupling between
the core and the envelope as the $M_{\rm c}-L_{\rm P}$ relation holds.

Actually, from the above discussion on the parameter $T_{\rm c}$, it
follows that the envelope structure is also quite insensitive to the
corresponding value of $R_{\rm c}$.  Nevertheless, it is a useful
exercise estimating $R_{\rm c}$ in an independent way and comparing it
with the predictions from Eq.~(\ref{t7r}).
 
Following the formalism of Tuchman et al.\ (1983), the core of an AGB
star behaves essentially as a {\it warm dwarf} of radius:
\begin{equation}
\label{rwd}
R_{\rm warm}( M_{\rm c},T) = k(M_{\rm c}, T) R_{\rm cold}(M_{\rm c})
\end{equation}
The term {\it warm dwarf} here indicates an isothermal
electron-degenerate stellar core with non-zero temperature.  The
radius, $R_{\rm cold}$, of a {\it cold dwarf} (defined at zero
temperature) is a function of its mass (and chemical composition)
only, whereas the radius of a {\it warm dwarf} is larger and depends
on the temperature too. The ratio $R_{\rm warm}/R_{\rm cold}$ is
expressed via the factor $k(M_{\rm c}, T)$ in Eq.~(\ref{rwd}), which
increases with $T$ and decreases with $M_{\rm c}$.  However, Tuchman
et al.\ (1983) pointed out that $k$ typically ranges from $1$ to $3$
for relevant burning shells and $M_{\rm c} \ga 0.6 \; M_{\odot}$.

The radius $R_{\rm cold}$ is calculated adopting the linear fit to the
Chandrasekhar (1939) results, as given by Tuchman et al.\ (1983):
\begin{equation}
\label{rcd}
R_{\rm cold} = 0.019 \, (1 - 0.58 M_{\rm c})
\end{equation}
where the quantities are expressed in solar units.

\begin{figure}
\resizebox{\hsize}{!}{\includegraphics{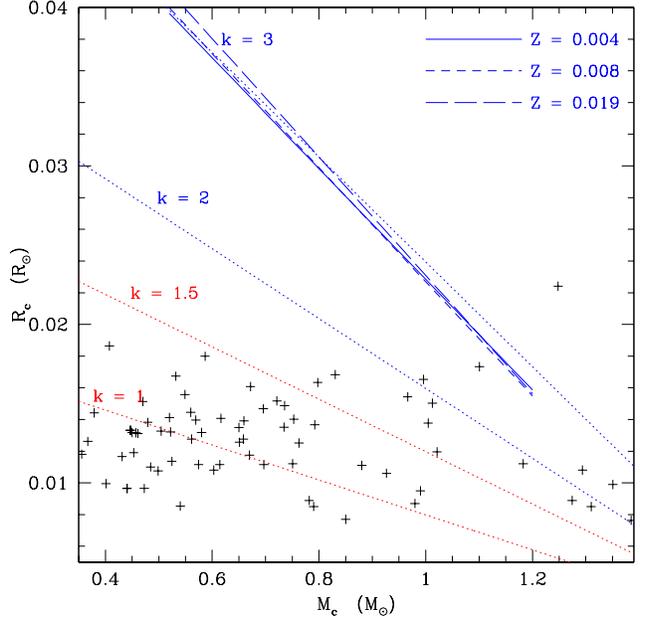}}
\caption{The adopted core radius -- core mass relationship for three
different values of the metallicity $Z$, derived from \protect
Eq.~(\ref{rcor}) with $T_{\rm c} = 10^{8}$ K.  The radius $R_{\rm
warm}$ calculated with Eq.~(\ref{rwd}) is shown for few values of $k$,
in the range $1 - 3$. The data corresponds to the empirical white
dwarf mass-radius relation (Koester 1987; Schmidt 1995; Koester \&
Reimers 1996).}
\label{fig_wdmr}
\end{figure}

Figure~\ref{fig_wdmr} shows $R_{\rm c}$ as function of $M_{\rm c}$,
calculated according to Eq.~(\ref{rcor}) with $T_{\rm c} = 10^{8}$ K
for three values of the envelope metallicity, $Z=0.004$, $Z=0.008$,
and $Z=0.019$. The radius $R_{\rm warm}$ as derived from
Eq.~(\ref{rwd}) for four different values of $k$ is also plotted.  For
such a choice of the parameter $T_{\rm c}$, the radius $R_{\rm c}$
turns out to be roughly $3 \times R_{\rm cold}$, in agreement with the
expected behaviour of {\it warm dwarfs}, and it is mostly comprised
within $0.04 \; R_{\odot}$ and $0.02 \; R_{\odot}$ for the relevant
range of $M_{\rm c}$ in AGB stars. Finally, for the sake of
comparison, the empirical data on the white dwarf mass-radius relation
(Koester 1987; Schmidt 1995; Koester \& Reimers 1996) is also
shown. It is worth noticing that the observed points are mostly
consistent with $1.5 \la k \la 1$.  This confirms the theoretical
expectation that white dwarfs approach the zero-temperature
configuration of complete degeneracy.

%%%%%%%%% %%%%%%%%%Figure %%%%%%%%
%%%%%%%%%%%%%%%%%%%%%%%%%%%%%%%%%%%

\subsubsection{The choice of ${T}_{\rm b}^{\rm dred}$}
\label{choice_of_t}

On the basis of the scheme just outlined, {\it the minimum core mass
for dredge-up of a model star of given total mass and surface chemical
composition is found (if it exists) when the condition of
Eq.~(\ref{cond_mcmin}) is satisfied for a unique pair of ($L_{\rm P},
T_{\rm eff}$) values}.

The method represents a considerable improvement since we can obtain
useful indications about the dependence of $M_{\rm c}^{\rm min}$ on
the envelope mass, metallicity, and peak strength, an important aspect
so far ignored in previous synthetic AGB calculations (see Groenewegen
\& de Jong 1993; Marigo et al.\ 1996a).

There is one more prescription to be discussed, i.e.\ the adopted
value of the characteristic temperature for convective dredge-up,
${T}_{\rm b}^{\rm dred}$. As already mentioned, indications come from
detailed analyses of thermal pulses, such as $\log {T}_{\rm b}^{\rm
dred} \sim 6.7$ in the work by Wood (1981), and $\log {T}_{\rm b}^{\rm
dred} \sim 6.5$ found by Boothroyd \& Sackmann (1988d).

Wood (1981) first performed a similar kind of analysis adopted here to
infer $M_{\rm c}^{\rm min}$ from envelope integrations. The author
aimed at investigating the theoretical incapability of reproducing the
observed carbon stars of low luminosity, i.e.\ no low-mass ($M < 3 \;
M_{\odot}$) and faint ($M_{\rm bol} > -5 $) carbon stars were
predicted to form according to full AGB calculations. He suggested
that the problem could be partially alleviated if the onset of the
dredge-up episodes occurs much earlier during the TP-AGB evolution of
low-mass stars. This implies a smaller $M_{\rm c}^{\rm min}$ than
found in full calculations (see Table~\ref{litmcmin}). The question
was considered by arbitrarily changing the value of the mixing-length
parameter, initially set to $\alpha \sim 1$. The net result was to
lower systematically the minimum total mass for dredge-up at
increasing $\alpha$, whereas the corresponding $M_{\rm c}^{\rm min}$
was slightly affected. Moreover, it became clear that, for given
stellar mass $M$, $M_{\rm c}^{\rm min}$ increases with $Z$ and, for
given $Z$, it decreases with $M$.
 
Boothroyd \& Sackmann (1988d) also addressed the problem of the
formation of low-mass carbon stars by analysing the effect of adopting
high values of the mixing length parameter [e.g.\ $\alpha \sim 3$ was
necessary to produce a $0.81 \; M_{\odot}, Z=0.002$ carbon star with
$M_{\rm c} \sim 0.566 \; M_{\odot}$ and a luminosity at the post-flash
dip $\log (L/L_\odot) \sim 3.34$, or equivalently $M_{\rm bol} \sim
-3.59$; see also Table~\ref{litmcmin}].  The authors remarked that the
value of $\alpha$ has a large effect on the depth in temperature of
the convective envelope, but it has a relatively small effect on the
depth in mass.  In fact, the base of the convective envelope is
located in a zone containing a very small mass [$ \sim (10^{-4} -
10^{-3}) M_{\odot}$], but across which the temperature gradient is
extremely steep.  Moreover, Boothroyd \& Sackmann (1988d) concluded
that the experiment of varying $\alpha$, though useful, cannot
definitively solve the problem of dredge-up.

These considerations suggest us the possibility of performing the
analysis on the occurrence of third dredge-up from a different
perspective, in the sense that we keep the mixing-length parameter
fixed at its value ($\alpha =1.68$) derived from the calibration of
the solar model, calculated with the complete evolutionary code of
Padua (Girardi et al.\ 1996).  
% addendum 
We specify that the adopted opacities are taken from Iglesias \&
Rogers (1996; OPAL) in the high-temperature domain ($T > 10^4$ K), and from
Alexander \& Feguson (1994) in the low-temperature domain ($T < 10^4$ K).
% addendum
We analyze, instead, the effect of
assuming different values of the characteristic temperature for
dredge-up. The lower $T_{\rm b}^{\rm dred}$, the smaller is the value
$M_{\rm c}^{\rm min}$. To this respect, we can safely state that
increasing $\alpha$ produces the same effect on $M_{\rm c}^{\rm min}$
as decreasing $T_{\rm b}^{\rm dred}$.

The reasons why we opt for such a choice are the following.  First, we
want to secure the continuity of the TP-AGB calculations with the
previous evolution, entirely followed using $\alpha =1.68$. This does
not mean at all that in reality the efficiency of the convective
mixing cannot change during the evolution of the star. However, since
the present theoretical treatment of the convection is indeed
inadequate to predict possible variations of $\alpha$, we prefer to
assume a conservative position. Second, the occurrence of convective
dredge-up is not only conditioned by the properties of external
convection (e.g.\ by the choice of $\alpha$), but it also depends on
the physical characteristics of the inter-shell region directly
involved in the flash. Detailed calculations clearly indicate that a
higher strength of the pulse, (usually measured by the maximum
luminosity of the He-shell) corresponds to a greater amount of energy
spent to expand outward the nuclearly processed material, that would
be cooled down more efficiently (Boothroyd \& Sackmann 1988d;
Straniero et al.\ 1997).

Therefore, artificially decreasing ${T}_{\rm b}^{\rm dred}$ is
equivalent to hypothesising stronger thermal pulses.  This is a viable
experiment in a theoretical approach making use of a static envelope
model. In fact, our analysis is not directed to investigate the
effective penetration of the convective envelope, since that would
require us to carefully follow (i.e.\ with a complete stellar code) the
onset and the time development of each thermal pulse.  Obviously, an
impossible task using only envelope integrations. The
essential goal of this study is to simply derive useful indications on
the possible occurrence of convective dredge-up, on the basis of
physical aspects which can be properly handled by our investigative
tool.

\subsubsection{The dependence of $M_{\rm c}^{\rm min}$ on $M$ and $Z$}
\label{results_for_mcmin}

%%%%%%%%% %%%%%%%%%Figure %%%%%%%%
\begin{figure}
\resizebox{\hsize}{!}{\includegraphics{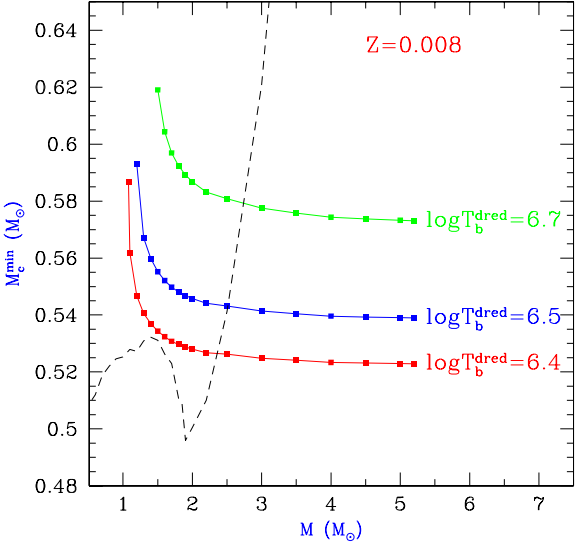}}
\caption{Minimum core mass for dredge-up, $M_{\rm c}^{\rm min}$, as a
function of the current stellar mass for models with [$Y=0.25,
Z=0.008$], corresponding to three values of ${T}_{\rm b}^{\rm dred}$.
The short-dashed line refers to the core mass at the first thermal
pulse.}
\label{mcminz008}
\end{figure}
%%%%%%%%%%%%%%%%%%%%%%%%%%%%%%%%%%%

%%%%%%%%% %%%%%%%%%Figure %%%%%%%%
\begin{figure}
\resizebox{\hsize}{!}{\includegraphics{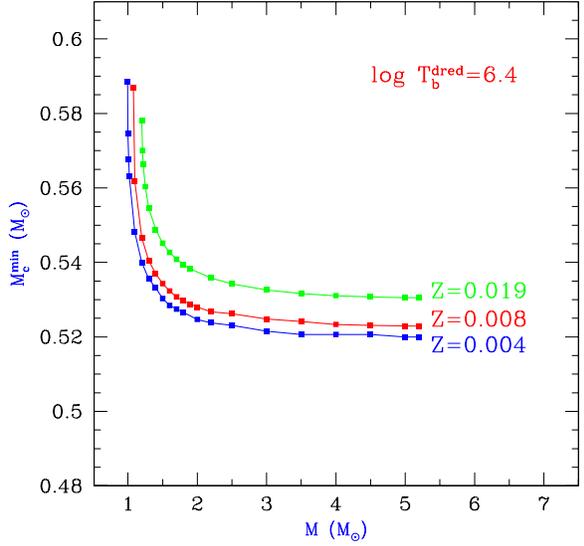}}
\caption{Minimum core mass for dredge-up, $M_{\rm c}^{\rm min}$, as a
function of the current stellar mass for models with metallicity
$Z=0.004, Z=0.008$, and $Z=0.019$, corresponding to the indicated
value of $\log T_{\rm b}^{\rm dred}$.}
\label{mcminz}
\end{figure}
%%%%%%%%%%%%%%%%%%%%%%%%%%%%%%%%%%%

In this section we aim at investigating the dependence of $M_{\rm
c}^{\rm min}$ on the stellar mass $M$ and metallicity $Z$, for
different values of the temperature parameter ${T}_{\rm b}^{\rm
dred}$.

As starting point, we explore both the cases $\log {T}_{\rm b}^{\rm
dred} = 6.7$ (from Wood 1981) and $\log {T}_{\rm b}^{\rm dred} = 6.5$
(following Booth\-royd \& Sackmann 1988d) to calculate $M_{\rm c}^{\rm
min}$ as a function of the stellar mass for the set of models with
original composition [$Y=0.250, Z=0.008$].  The results are
displayed in Fig.~\ref{mcminz008}, together with the case $\log
{T}_{\rm b}^{\rm dred} = 6.4$.

One can immediately notice that all three curves share a common trend,
i.e.\ $M_{\rm c}^{\rm min}$ very steeply increases toward lower masses
[$M < (1.5 - 1.8) M_{\odot}$], and flattens out to a nearly
constant value for greater masses ($M > 2 M_{\odot}$).  The comparison
of these curves with the values of the core mass at the starting of
the thermally pulsing regime (short-dashed line) allows one to get an
approximate estimate of the core-mass increment necessary before the
star begins to become carbon-enriched. It turns out that stars with
masses $M > 2.5 M_{\odot}$ enter the TP-AGB phase with a core mass
that is greater than the corresponding $M_{\rm c}^{\rm min}$, for all
the three values of temperature ${T}_{\rm b}^{\rm dred}$ under study,
i.e.\ the third dredge-up should already occur at the first pulse (or,
more likely, as soon as the full-amplitude regime is established).  On
the contrary, stars with lower masses ($M < 2.5 M_{\odot}$) are
expected to evolve through a longer part of the TP-AGB without
undergoing any chemical pollution by convective dredge-up.

Moreover, it is worth noticing that each possible choice of ${T}_{\rm
b}^{\rm dred}$ determines a different value of the minimum total mass,
$M_{\rm dred}^{\rm min}$, for a star to experience dredge-up. In the
$Z=0.008$ case, calculations yield $M_{\rm dred}^{\rm min} \sim 1.5
M_{\odot}$ adopting $\log {T}_{\rm b}^{\rm dred} = 6.7$, $M_{\rm
dred}^{\rm min} \sim 1.2 M_{\odot}$ adopting $\log {T}_{\rm b}^{\rm
dred} = 6.5$, and $M_{\rm dred}^{\rm min} \sim 1.1 M_{\odot}$ adopting
$\log {T}_{\rm b}^{\rm dred} = 6.4$.

The dependence on metallicity is illustrated in Fig.~\ref{mcminz}, for
a fixed value of $\log T_{\rm b}^{\rm dred}=6.4$.  In agreement with
the results from full computations of thermal pulses (e.g.\ Boothroyd
\& Sackmann 1988d), our analysis confirms that for given stellar mass,
the onset of convective dredge-up is favoured at lower metal
abundances in the envelope. It is worth noticing that the minimum
mass, $M_{\rm dred}^{\rm min}$, is affected in the same
direction. According to our calculations, we get $M_{\rm dred}^{\rm
min} \sim 1.2 M_{\odot}$ for $Z=0.019$, $M_{\rm dred}^{\rm min} \sim
1.1 M_{\odot}$ for $Z=0.008$, and $M_{\rm dred}^{\rm min} \sim 1
M_{\odot}$ for $Z=0.004$.  This result seems to be consistent with the
observed data of carbon stars in the Magellanic Clouds (see
Sect.~\ref{cslf}): the faint end of the luminosity distribution of
carbon stars in the SMC (with a mean metallicity $Z \sim 0.004$)
extends to a fainter luminosity ($M_{\rm bol} \sim -2.5$), than in the
LMC ($M_{\rm bol} \sim -3$, with a mean $Z \sim 0.008$), thus
suggesting a lower $M_{\rm dred}^{\rm min}$ at lower metallicities.

We like to stress once more that the onset of the mixing events in
low-mass stars is a fundamental aspect, since it crucially affects the
distribution of the carbon stars in the low-luminosity domain. As will
be shown below (from a systematic analysis), the value $\log {T}_{\rm
b}^{\rm dred} =6.4$ is found suitable to fulfill the above constraint
(Sect.~\ref{cslf}) in the Magellanic Clouds.  

It is worth specifying that the results illustrated in
Fig.~\ref{mcminz008} give $M_{\rm c}^{\rm min}$ as a function of the
{\em current total mass} (not the mass at the first thermal
pulse). This is particularly relevant in the low-mass domain, where
the curves become very steep, i.e.\ minor variations of the total mass
correspond to quite different values of $M_{\rm c}^{\rm min}$.  Since
TP-AGB calculations are performed with the inclusion of mass loss by
stellar winds, it follows that {\em in the case of low-mass stars a
small reduction of the envelope mass may significantly delay, or even
prevent the onset of dredge-up}.

%%%%%%%%% %%%%%%%%%Figure %%%%%%%%
\begin{figure}
\resizebox{\hsize}{!}{\includegraphics{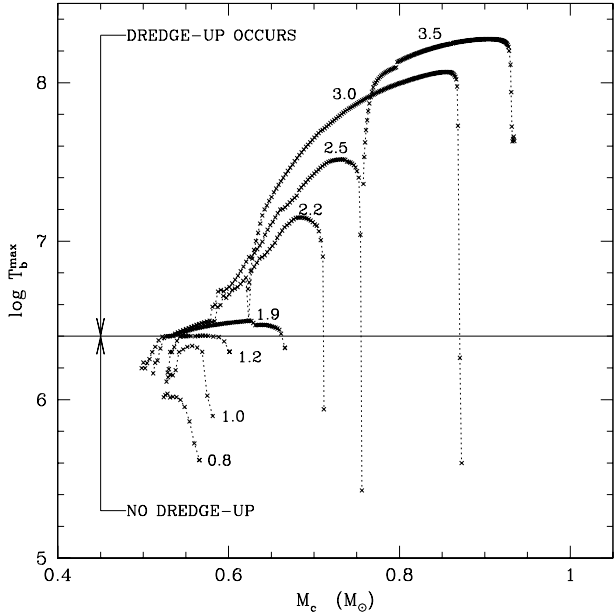}}
\caption{Maximum base temperature, $T_{\rm b}^{\rm max}$, calculated
at each thermal pulse (crosses connected by dotted lines) as a
function of the core mass for a few evolving TP-AGB stellar models
with original metallicity, $Z=0.008$. Stellar masses 
at the onset of the TP-AGB phase are indicated in
solar units nearby the corresponding curves.  The solid horizontal
line  marks the critical
temperature ($\log T_{\rm b}^{\rm dred} = 6.4$) for the
occurrence of the third dredge-up. }
\label{ftdred}
\end{figure}
%%%%%%%%%%%%%%%%%%%%%%%%%%%%%%%%%%%

Figure~\ref{ftdred} shows the maximum temperature at the base of the
convective envelope, $T_{\rm b}^{\rm max}$, at the stage of post-flash
luminosity peak, as a function of the core mass during the evolution
of a few stellar models with different masses $M$ at the beginning
of the TP-AGB phase, and
original metallicity $Z=0.008$ (see Sect.~\ref{mods} for further
details of the models).  The adopted temperature parameter is
${T}_{\rm b}^{\rm dred}=6.4$.

Envelope integrations indicate that if the envelope mass is too small,
the star fails to attain the minimum temperature ${T}_{\rm b}^{\rm
dred}$, whatever the core mass is. See, for instance, the track for
the $M = 0.8 M_{\odot}$ star.  The model with $M = 1.2 M_{\odot}$
corresponds to the lowest mass for a star to become
carbon-enriched.  Higher mass stars are able to attain hot base
temperatures, so that they are expected to experience dredge-up
already since the first pulses.

The present method also provides the possibility of predicting the
stage beyond which dredge-up cannot occur any longer during the final
TP-AGB evolution. The shut-down of dredge-up may happen when the
envelope mass becomes so small that the condition $T_{\rm b}^{\rm max}
\ge {T}_{\rm b}^{\rm dred}$ is no longer satisfied. This circumstance
occurs for the last pulses of the tracks displayed in
Fig.~\ref{ftdred}, characterized by the drastic reduction of the
envelope mass due to stellar winds.  This point represents another
important improvement for synthetic TP-AGB models, since the end of
the dredge-up episodes is not fixed a priori (a necessary assumption
if one adopts a constant value for $M_{\rm c}^{\rm min}$), but
naturally derives from calculations.

%%%%%%%%%%%%%%%%%%%%%%%%%%%%%%%%%%%%%%%%%%%%%%%%%%%%%%%%%%%%%%%%%
\subsection{The inter-shell chemical composition}
\label{chemdred}

The remaining ingredient of our models is the composition of the
dredged-up material. We use the same formalism as in Marigo et al.\
(1996a), to whom we refer for an extensive description of this point.
Suffice it to recall that our prescription is based on the detailed
calculations of the inter-shell nucleosynthesis carried out by
Boothroyd \& Sackmann (1988c). They find that the convective
inter-shell after a thermal pulse and just before the penetration of
the envelope is composed by $20-25$~\% of $^{12}$C, $\sim2$~\% of
$^{16}$O, and the remainder of $^{4}$He (in mass fractions).  These
proportions are reached after the first few thermal pulses, and do not
significantly depend on the core mass and metallicity.

Recently, Herwig et al.\ (1997) have pointed out that the abundances
in the inter-shell will differ significantly from the 
results of standard calculations
(i.e.\ without extra-mixing as in Boothroyd \& Sackmann 1988c), if
instead one applies their overshooting scheme to all convective zones.
The resulting chemical distribution of the dredged-up material
typically consists of $50 \%$ in $^{12}$C, $25 \%$ in $^{16}$O, and
$25 \%$ in $^{4}$He. 

In this study we opt for the conservative standard prescription of
Boothroyd \& Sackmann (1988c). The following fixed abundances are
adopted: 22\% for $^{12}$C produced by the triple alpha reaction and
2\% for $^{16}$O produced by the reaction $^{12}$C($\alpha,
\gamma$)$^{16}$O. 

%%%%%%%%%%%%%%%%%%%%%
\subsection{Other computational details}
\label{mods}
The evolutionary calculations presented in this study have been
carried out for a dense grid of stellar masses ($0.8 M_{\odot} \la \mi
\le 5 M_{\odot}$) and two values of the initial metallicity, $Z=0.008$
and $Z=0.004$.  In addition to the new treatment of the third
dredge-up already presented, the present calculations are based on a
synthetic TP-AGB model, having the same general structure as described
in Marigo et al.\ (1996a), with some major revisions and updates
illustrated in Marigo et al.\ (1998) and Marigo (1998ab). The reader
should refer to these works for all the details. Suffice it to recall
here that the synthetic model combines analytical relations with a
complete stellar envelope, which is integrated from the photosphere
down to the core.  The evolution of a TP-AGB star is followed from the
first thermal pulse up to the complete ejection of the envelope,
adopting the prescription for mass loss as suggested by Vassiliadis \&
Wood (1993).  A homogeneous set of accurate analytical relations
(e.g.\ the core mass-luminosity relation, the core mass-interpulse
period) is taken from the recent work by Wagenhuber \& Groenewegen
(1998).  Finally, it is worth remarking that the occurrence of
envelope burning (or hot-bottom burning) in more massive TP-AGB stars
($M \ga 3.5 M_{\odot}$) is consistently followed to account for the
possible break-down of the $M_{\rm c}-L$ relation (see Marigo 1998b).

%%%%%%%%%%%%%%%%%%%%%%%%%%%%%%%%%%%%%%%%%%%%%%%%%%%%%%%%%%%%%%%%%%%%%%%%
\section{The observed carbon star luminosity functions in the MCs}
\label{cslf}

The Magellanic Clouds -- LMC and SMC -- are an ideal site for testing
evolutionary theories of AGB stars, thanks to the fortunate
combination of several factors:
\begin{itemize}
\item The extinction to both Clouds is low and relatively uniform.
\item The galaxies are far enough ($\sim 50$ kpc to the LMC, $\sim 63$
kpc to the SMC) that their depth along the line of sight can be
neglected to first order, and the same distance can be assigned to all
the stars.
\item At the same time the galaxies are near enough to allow
the resolution and analysis of single stellar sources by
means of both photometric and spectroscopic techniques.
\item Both Clouds have a considerable number of populous star clusters
whose age (Elson \& Fall 1985; Girardi et al.\ 1995) and chemical
composition (Olszewski et al.\ 1991) can be estimated with a good
accuracy, and which contain a considerable number of AGB stars (Frogel
et al.\ 1990; Westerlund et al.\ 1991).
\item A large amount of observed data on AGB stars is nowadays
available (see the reviews by Westerlund 1990 and Groenewegen 1997),
thanks to large scale surveys ranging from the optical to the
radio domain, and spectroscopic analyses for the estimation of stellar
parameters (e.g.\ effective temperature, mass-loss rate, wind velocity,
surface gravity, surface chemical abundances).
\end{itemize}    

%%%%%%%%%%%%%% FIGURE %%%%%%%%%%%%%%%%%%%%%%%%%%%%%%%
%
\begin{figure}
\resizebox{\hsize}{!}{\includegraphics{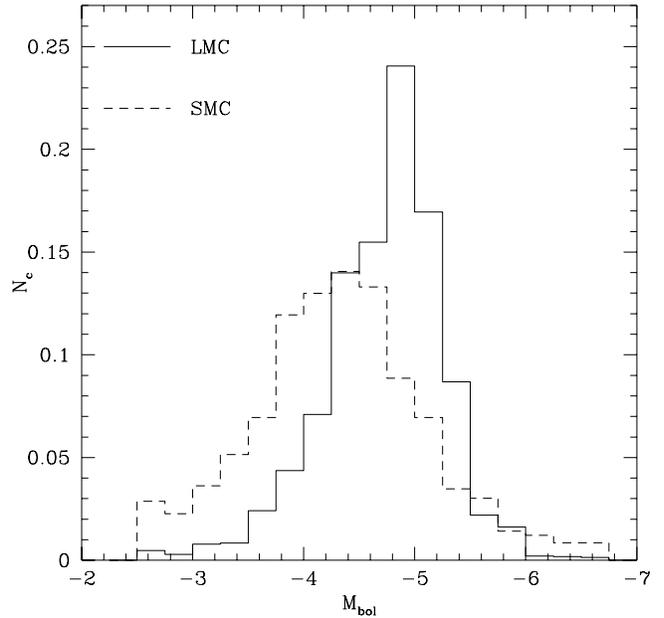}}
\caption{The observed LFs of carbon stars detected in the fields of
the LMC (solid line) and SMC (dashed line), normalised to the total
number of stars in the samples.  The data for the LMC are taken from
Costa \& Frogel (1996), adopting a distance modulus $\mu_0 = 18.5$.
The data for the SMC is as derived by Groenewegen (1997) from the data
compiled by Rebeirot et al.\ (1993), with a distance modulus $\mu_0 =
19.0$.}
\label{cstarlmc}
\end{figure}
%
%%%%%%%%%%%%%%%%%%%%%%%%%%%%%%%%%%%%%%%%%%%%%%%%%%%%%%%%%%%

For the observed carbon star luminosity function in the LMC we make
use of the recent compilation of Costa \& Frogel (1996).  The data
refer to $895$ carbon stars, selected among the original sample of
$1035$ stars found by Blanco et al.\ (1980) and Blanco \& McCarty
(1983) by means of low-dispersion red grism spectroscopy in $52$
fields of the galaxy, distributed over an area of $8^{\circ} \times
8^{\circ}$.  The identification of carbon stars was based on the
recognition in stellar spectra of a typical blend of the CN bands at
near-infrared wavelengths ($\lambda \lambda 7910, 8100, 8320$).  It is
worth remarking that this sample of carbon stars can be considered
almost complete and not limited by apparent magnitude, since the
faintest carbon stars are at least 2~mag brighter ($I \sim 15$)
than the faint limit of detection ($I \sim 17$).  Moreover, the
possible contribution of high luminosity carbon stars, which would
have been missed in the survey because of dust obscuration, is likely
to be rather small ($\la 3 \%$ for $M_{\rm bol} < -6$ according to
Groenewegen \& de Jong 1993).  Costa \& Frogel (1996) performed $RI$
and/or $JHL$ photometry of the selected $895$ carbon stars. The
apparent bolometric magnitudes $m_{\rm bol}$ were estimated first
applying proper bolometric corrections to the $K$ magnitude of a
subsample of stars with available $JHK$ photometry.  On the basis of
these results the authors derived a linear relationship, expressing
($I_0 - m_{\rm bol}$) as a function of the red colour $(R-I)_0$, which
was then used to obtain $m_{\rm bol}$ for the majority of carbon stars
with only $RI$ data.  The authors quoted a mean error on the
evaluation of $m_{\rm bol}$ of approximately $\pm 0.34$ mag.

The observed CSLF in the LMC is plotted in Fig.~\ref{cstarlmc} (solid
line).  The absolute bolometric magnitudes, $M_{\rm bol}$, are
obtained from the apparent ones adopting a distance modulus $\mu_0 =
m_{\rm bol}-M_{\rm bol} = 18.5$, as indicated by Westerlund (1990) and
favoured by several recent works (e.g.\ Wood et al.\ 1997; Oudmaijer et
al.\ 1998; Madore \& Freedman 1998; Salaris \& Cassisi 1998; Panagia
1998).

The resulting CSLF is quite similar to those obtained from previous
samples of carbon stars in the fields of the LMC (Westerlund et al.\
1978; Richer et al.\ 1979; Cohen et al.\ 1981).  It extends
approximately from $M_{\rm bol}=-3$ up to $M_{\rm bol}=-6.5$, with the
peak located at around $M_{\rm bol}=-4.875$ (centre of the bin).  The
dispersion of the observed distribution is estimated to be about
$\sigma = 0.55$~mag.

From a comparison with the CSLF in selected clusters of the LMC
(Frogel et al.\ 1990) it turns out that the two distributions are
quite similar, except for few subtle differences, namely, the presence
of somewhat more extended faint and bright tails in the field
distribution relative to that of the clusters.  According to Marigo et
al.\ (1996b), the explanation of this resides in the presence of two
different periods of quiescence in the cluster formation history of
the LMC, shaping the age (and progenitor mass) distribution of C
stars.  These gaps are found at ages of $\sim (3 - 12) \times
10^{9}$ yr, and $\sim (2 - 6) \times 10^{8}$ yr (see also Girardi
et al.\ 1995), which roughly correspond to the ages of the least
($\sim 1.2 M_{\odot}$) and most ($\sim 4 M_{\odot}$) massive
carbon star progenitors, respectively.

Figure~\ref{cstarlmc} also shows the observed CSLF in the SMC (dashed
line), as derived by Groenewegen (1997) from a sample of $1636$ stars
observed by Rebeirot et al.\ (1993). The adopted bolometric correction
are those by Westerlund et al.\ (1986).  The adopted distance modulus
for the SMC is $\mu_{0} = 19.0$ (see Westerlund 1990 for a review of
SMC distance determinations).

The first bin of this CSLF contains all stars fainter than $M_{\rm
bol} = -2.5$. They are less than 3~\% of the total sample of SMC
carbon stars. Their intrinsic magnitudes can be as low as $M_{\rm bol}
= -1.8$, as found by Westerlund et al.\ (1995). The formation of
carbon stars at such low luminosities can hardly be understood as the
result of the third dredge-up in single AGB stars. Most probably, they
arise from the evolution in close binary systems (see e.g.\ de Kool \&
Green 1995; Frantsman 1997), in which the surface of a low-mass star
has been contaminated by the ejecta of a former AGB star
companion. The low-luminosity carbon stars will not be considered in
this work, since we are dealing with single star evolution. For all
practical comparisons, we will consider the CSLF in the SMC as
extending down to $M_{\rm bol} = -2.5$.

The CSLFs are significantly different in the two Magellanic Clouds.
In the SMC the CSLF is much broader, and presents a fainter peak
location (at $M_{\rm bol} \sim -4.375$; centre of the bin) than that
in the LMC ($M_{\rm bol} \sim -4.875$).  This indicates that LMC
carbon stars are, on the mean, about 60~\% brighter than their SMC
counterparts. As remarked by Groenewegen (1997), explaining this
difference quantitatively represents a challenge for synthetic TP-AGB
models.

%%%%%%%%%%%%%%%%%%%%%%%%%%%%%%%%%%%%%%%%%%%%%%%%%%%%%%%%%%%%%%%%%
\section{The theoretical carbon star luminosity function}
\label{theo_cslf}

In order to simulate the observed CSLF in the Magellanic Clouds, we
employ the general scheme already used in Marigo et al.\ (1996a). For
the sake of completeness, we here simply recall the basic steps.

Dividing the luminosity range of interest (e.g.\ $-2.5 > M_{\rm bol} >
-7.0$) into a suitable number of intervals, for each ${k^{\rm th}}$
luminosity bin of width $\Delta M_{\rm bol}$, we are required to
evaluate the number $N_{k}$ of carbon stars which are expected to be
in transit at the present epoch:
\begin{eqnarray} 
N_{k} & = &
\int_{M_{\rm i}^{\rm min}}^{M_{\rm i}^{\rm max}} 
	N_{k}(\mi) \diff\mi 
	\nonumber \\
& \propto &
\int_{M_{\rm i}^{\rm min}}^{M_{\rm i}^{\rm max}} 
	\Phi (\mi, t) \Delta \tau_{k}(\mi) \diff\mi
	\label{dncstar}
\end{eqnarray}
where $\Phi(\mi, t)$ is the stellar birth rate, i.e.\ the number of
stars formed with mass \mi\ at the epoch $t$, per unit of mass and
time (see Sect.~\ref{sbr}); and $\Delta \tau_{k}(\mi)$ is the time
each star of mass $\mi$ spends in the $k^{\rm th}$ luminosity
interval.  Thus, $N_{k}(\mi)$ expresses the number of carbon stars
evolved from a progenitor with initial mass \mi\ and a current (i.e.\
at the present time) luminosity inside the $k^{\rm th}$ bin.

The integration is performed over the mass range $[\mi^{\rm min}$,
$\mi^{\rm max}]$. We denote with $\mi^{\rm min}$ the maximum of
the lower mass limit for the formation of carbon stars and the mass of
the star whose lifetime is just the age of the galaxy ($\sim 0.87
M_{\odot}$ for $T_{\rm G} =15$ Gyr). We refer to $\mi^{\rm max}$ as
the minimum of the upper mass limit for carbon star formation,
and the maximum mass of stars which experience the AGB phase, $M_{\rm
up}$. In our case, the evolution of the progenitor stars has been
calculated with the adoption of an overshoot scheme applied to the
convective zones, from the ZAMS up to the starting of the TP-AGB
(Girardi et al.\ 1998). The overshooting parameter for core convection
adopted for intermediate-mass stars ($\Lambda_{\rm c}=0.5$)
characterises a moderate efficiency for this process. This fixes
$M_{\rm up} \simeq 5 M_{\odot}$.  Classical models without overshoot
would predict a higher value, e.g.\ $M_{\rm up} \sim (7 - 8)
M_{\odot}$.

Let us now discuss the meaning of the two basic functions in
Eq.~(\ref{dncstar}), namely $\Phi (\mi,t)$ and $\Delta \tau_{k}(\mi)$.

\subsection{The stellar birth rate}
\label{sbr}

It is useful to split the stellar birth rate $\Phi$ into the product
of two functions (see Tinsley 1980):
\begin{equation}
\Phi(\mi,t) = \phi(\mi) \, \psi(t)
\end{equation}
where $\phi(\mi)$ is the initial mass function (IMF) expressing the
fraction of stars forming with masses in the range $[\mi,
\mi+\diff\mi]$. This latter is normalised so that:
\begin{equation}
\int_0^{\infty} \mi\, \phi(\mi)\, \diff\mi = 1
\end{equation}
In this work we adopt the classical power-law IMF of Salpeter (1955),
$\phi(\mi) \propto \mi^{-(1+x)}$, with $x=1.35$.  The function
$\psi(t)$ is the star formation rate (SFR) and represents the mass of
primordial gas that is converted into stars per unit of time at the
epoch $t$.

In the particular case under consideration [see Eq.~(\ref{dncstar})],
it is required to evaluate the number of carbon stars with initial
mass $\mi$ and lifetime $\tau^*$ which are born at the epoch $t = T_{\rm
G} - \tau^*$ (and hence are presently observable), 
where $T_{\rm G}$ is the age of the galaxy.  This can be expressed by:
\begin{equation}
\Phi(\mi,T_{\rm G}-\tau^*) = \phi(\mi)\,\psi[T_{\rm G} - \tau^*(\mi)]
\label{SFF}
\end{equation}
Since there is a unique correspondence between the lifetime of a star
(of given metallicity) and its initial mass, $\tau^* =\tau^*(\mi)$, it
is possible to consider $\Phi$ as a function solely of the stellar
mass.

\subsection{The rate of brightening}
\label{ratebri}

The quantity $\Delta \tau_{k}(\mi)$ is the time that a star with
initial mass $\mi$ spends in the $k^{\rm th}$ luminosity bin of width
$\Delta M_{\rm bol}$. It is directly derived from the TP-AGB
evolutionary tracks.

A useful approximation for interpreting the 
results is that the quiescent luminosity of TP-AGB stars
follows an almost constant rate of brightening (see e.g.\ Iben 1981):
\begin{equation}
\frac{\diff M_{\rm bol}}{\diff\tau}  \sim {\rm constant}
\label{constant_rate}
\end{equation}
Actually, this relation holds if the following conditions are met.
First, the relation between the quiescent luminosity $L$ and the core
mass $M_{\rm c}$ should be linear.  This would correspond to consider
only the terms (5a) and (5b) of Eq.~(5) giving the $M_{\rm c}-L$
relation adopted by Marigo (1998b). This implies the neglect of
both the initial sub-luminous evolution of the first inter-pulse
periods, and the possible deviations towards higher luminosities
caused by envelope burning.  This assumptions are reasonable for
carbon stars since their formation generally occurs after the onset of
the full amplitude regime, and only when envelope burning is not
efficiently operating. In this case:
\begin{equation}
\frac{\diff M_{\rm c}}{\diff L} = K_{1} 
\end{equation}
On the other hand, the evolutionary rate (i.e.\ the growth rate of the
core mass) is proportional to the stellar luminosity:
\begin{equation}
\frac{\diff M_{\rm c}}{\diff \tau} = K_{2} \frac{L}{X} 
\label{mcrate}
\end{equation}
In the above equations $K_1$ and $K_2$ are suitable constants.  If the
changes in the envelope abundance of hydrogen, $X$, are small enough,
it follows that $\diff M_{\rm c}$ can be eliminated from the above
equations giving $d\ln L/d\tau={\rm constant}$, which is equivalent to
Eq.~(\ref{constant_rate}).  This equation tells us that the {\it rate of
brightening} along the TP-AGB is nearly constant, regardless of the
core mass, the total mass, and the luminosity level.

In our computations, the rate of brightening for carbon stars is found
to be constant within $\sim$ 35~\% at most.  Typical values for this
quantity are $\sim -(3.5 - 7)\;10^{-7}$mag\,yr$^{-1}$.  Under these
conditions, it follows that nearly the same number of stars is
expected to populate each magnitude interval along the AGB phase of
coeval (e.g.\ a cluster) stellar populations. If we consider only the
evolution of the quiescent luminosity (see, however, Sect.~\ref{dip}
below), the resulting CSLF would be described by a box-like
distribution.

A final remark is worth making. Since dredge-up periodically reduces the core
mass, then it is expected to affect the average evolutionary rate
$\diff M_{\rm c}/\diff \tau$.  This implies that Eq.~(\ref{mcrate}) is
valid only if either dredge-up is not operating, or $\lambda={\rm
constant}$ as in our prescriptions. In contrast, for $\lambda$
variable with time, neither $K_2$ nor the rate of
brightening $\diff M_{\rm bol}/\diff \tau$ would be constant.

\subsection{Transition luminosities}
\label{bolcrit}
 
The calculation of the CSLF requires also the knowledge of:
\begin{itemize}
\item the luminosities marking the transition from the
M-class (C/O $< 1$) to the C-class (C/O $> 1$),
\item the maximum luminosities at which the star is still a carbon
star,
\end{itemize}
as a function of the initial mass.  

%%%%%%%%%%%%%%%%% FIGURE  2 %%%%%%%%%%%%%%%%%%%%%%%%%%%%%%%%%%%%%%%%
\begin{figure}
\resizebox{\hsize}{!}{\includegraphics{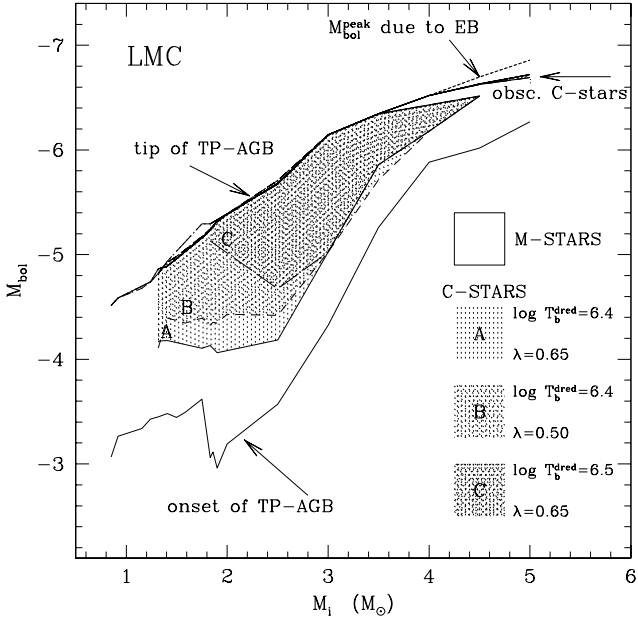}}
\caption{Transition bolometric magnitudes vs.\ initial masses of TP-AGB
stars with original metallicity $Z = 0.008$. Regions labelled with
letters A, B and C refer to three choices of the dredge-up parameters.
See the text for more details.}
\label{limbolz008}
\end{figure}
%%%%%%%%%%%%%%%%%%%%%%%%%%%%%%%%%%%

%%%%%%%%%%%%%%%%% FIGURE  2 %%%%%%%%%%%%%%%%%%%%%%%%%%%%%%%%%%%%%%%%
\begin{figure}
\resizebox{\hsize}{!}{\includegraphics{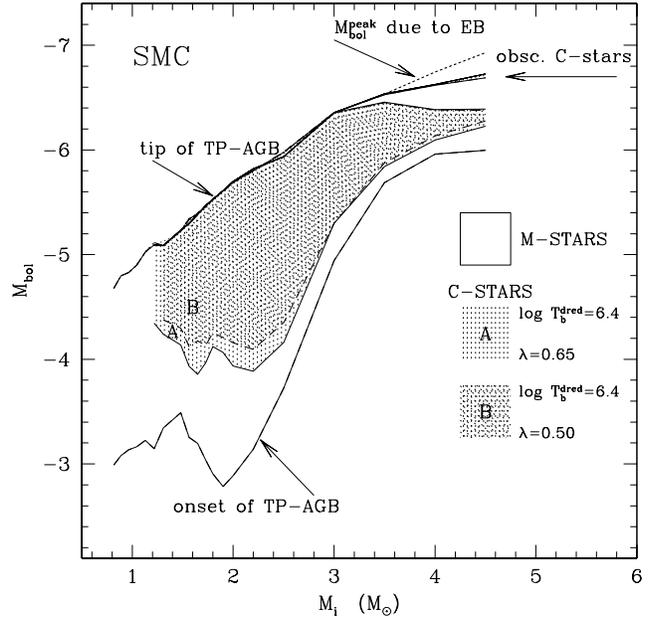}}
\caption{Transition bolometric magnitudes vs.\ initial masses of
TP-AGB stars with original metallicity $Z = 0.004$.  See the text for
more details.}
\label{limbolz004}
\end{figure}
%%%%%%%%%%%%%%%%%%%%%%%%%%%%%%%%%%%

Figures \ref{limbolz008} and \ref{limbolz004} show the transition
luminosities as a function of the initial mass for the $Z = 0.008$ and
$Z = 0.004$ cases, respectively. The choice of dredge-up parameters
adopted to construct these plots will be discussed later in
Sects.~\ref{calibr} and ~\ref{smc}.  Several interesting features can
be noticed from the inspection of these figures.

First, the formation of carbon stars may be prevented because of
missing or insufficient dredge-up in the range of lower masses, or
significantly delayed due to the occurrence of envelope burning in the
most massive AGB stars.

Second, for the majority of low-mass carbon stars (with no envelope
burning; $\mi \la 3.5 M_{\odot}$; see Marigo 1998b) the maximum
luminosity just coincides with the tip of the AGB reached at the end
of their evolution. On the contrary, in more massive carbon stars
($\mi \ga 3.5 M_{\odot}$) the occurrence of envelope burning can lead
to a different configuration.  It may happen that the onset of
envelope burning takes place after the star has already experienced a
sufficient number of dredge-up episodes to become a carbon star.
Nuclear burning at the base of the envelope subsequently destroys
carbon, lowering the C/O ratio again below unity.  Then, the star is
expected to evolve like an oxygen-rich star, until a new transition to
the C-class possibly occurs in the very late stages of the
evolution. In fact, after envelope burning has extinguished, the star
eventually undergoes further convective dredge-up of carbon before
leaving the AGB phase (see Frost et al.\ 1998).  Then the star would
appear as a carbon star in two distinct evolutionary stages, being
oxygen-rich in between. This latter configuration can be clearly
noticed in Figs.~\ref{limbolz008} and \ref{limbolz004}: at masses
higher than about $3.5 M_{\odot}$, a thin strip indicates the domain
of the high-luminosity carbon stars, i.e.\ those which experienced a
late conversion to this spectral type.

In this case, the transition to the carbon star phase may occur under
conditions of severe mass-loss. Thus, the brightest carbon stars are
expected to be obscured by thick layers of ejected material, then not
detectable at visual wavelengthes. Actually, high-luminosity obscured
carbon stars have been recently discovered in the LMC, by means of
infra-red observations (van Loon et al.\ 1998). Since they are not
included in the observed CSLFs in the Magellanic Clouds
(Sect.~\ref{cslf}), we will not consider them here.  The formation of
these carbon stars will be investigated by means of the present models
in a forthcoming paper.

\subsection{The effect of the low-luminosity dip}
\label{dip}

The fact that low-mass ($\mi \la 3 M_{\odot}$) TP-AGB stars undergo a
long-lived subluminous stage, soon after the occurrence of a thermal
pulse, plays a crucial role in determining the extension of the
low-luminosity tail of the CSLF (Boothroyd \& Sackmann 1988ad).  It
turns out that without accounting for this effect the theoretical
distribution cannot extend down to the observed faint end (see also
Groenewegen \& de Jong 1993), unless quite extreme (and unlikely)
assumptions are adopted, e.g.\ the occurrence of very efficient
($\lambda \sim 1$) dredge-up events since the first weak thermal
pulses. On the contrary, the effect of the post-flash luminosity
maximum on the CSLF can be neglected, due to its very short duration
(Boothroyd \& Sackmann 1988a).

In our calculations, the drop in luminosity below the $M_{\rm c} - L$
relation is included in a very simple way. On the basis of the results
from complete calculations of thermal pulses (Boothroyd \& Sackmann
1988a) we construct a function, $P(L-\Delta L)$, expressing the
probability that a star with quiescent luminosity $L$, predicted by
the $M_{\rm c}-L$ relation, may be found at a lower luminosity level,
$L-\Delta L$.  The maximum excursion towards fainter luminosities is
assumed to be $\Delta \log L = 0.5$, corresponding to $\Delta M_{\rm
bol} = 1.25$ mag.  The probability function is then estimated for a
suitable number of luminosity intervals (5 in our case), with a width
equal to the adopted bin resolution, i.e.\ $\Delta \log L = 0.1$, or
equivalently $\Delta M_{{\rm bol}} = 0.25$.

The probability distribution, normalized to unity, is assumed as
following:
\begin{eqnarray}
P_0(\log L) \,\,\,\,\, & = & 0.35 \nonumber \\
P_1(\log L-0.1) & = & 0.28 \nonumber \\
P_2(\log L-0.2) & = & 0.20 \nonumber \\
P_3(\log L-0.3) & = & 0.10 \\
P_4(\log L-0.4) & = & 0.05  \nonumber \\
P_5(\log L-0.5) & = & 0.02  \nonumber
\label{p_dip}
\end{eqnarray} 
Then, the number $N_{k}$ of carbon stars [see Eq.~(\ref{dncstar})]
predicted to populate the $k^{\rm th}$ luminosity bin is calculated
according to:
\begin{equation}
N_{k} = 
N_{k,0} \, P_{0} + 
\sum_{i=1}^{5} N_{k+i} \, P_{i}
\end{equation}
where $N_{k,0} \, P_{0}$ is the number of objects, following the
$M_{\rm c}-L$ relation, which are expected to be presently crossing
the $k^{\rm th}$ interval.  The term $N_{k+i} \, P_{i}$ represents
the number of stars having higher quiescent luminosity within the
$(k+i)^{\rm th}$ bin, but presently contributing to the $k^{\rm th}$
bin, because of the overlap with the corresponding low-luminosity dip.
In other words, the luminosity function derived just applying the
$M_{\rm c}-L$ relation, is then convolved with the function $P(L)$
which mimics the shape of the sub-luminous stages driven by thermal
pulses.

We have actually adopted the same  prescription for all the stars
considered, although the duration and the depth of the post-flash
luminosity dip are less pronounced in stars of higher core masses
(Boothroyd \& Sackmann 1988a).  Nevertheless, we expect that our
approximation does not compromise the validity of the results. This
because most of the carbon stars are found in the mass range $\mi <
2.5 M_{\odot}$, for which well extended luminosity dips, consistent
with Eq.~(\ref{p_dip}), are expected. Moreover, it will be shown in
Sect.~\ref{calibr} that these stars completely determine the faint end
and peak location of the theoretical CSLFs. More massive AGB stars,
instead, are predicted to have less-pronounced luminosity
dips. However, they populate mostly the bright wing of the CSLF, which
is affected by several other factors (such as the efficiency of
envelope burning, mass-loss, and the SFR; see Sect.~\ref{calibr}).

%%%%%%%%%%%%%% FIGURE 4  %%%%%%%%%%%%%%%%%%%%%%%%%%%%%%%
%
\begin{figure}
\resizebox{\hsize}{!}{\includegraphics{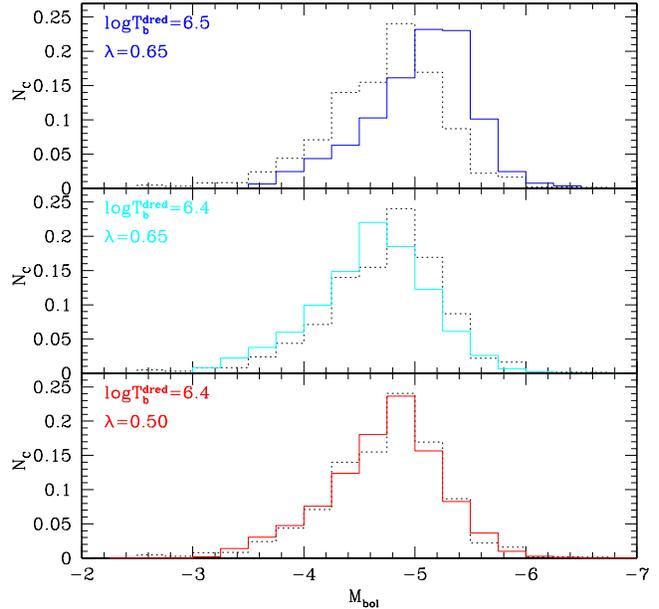}}
\caption{Theoretical luminosity functions of carbon stars (solid
lines) for three different combinations of the dredge-up parameters as
indicated.  The observed distribution in the LMC (dotted line) is
plotted for comparison.}
\label{cstar3}
\end{figure}
%
%%%%%%%%%%%%%%%%%%%%%%%%%%%%%%%%%%%%%%%%%%%%%%%%%%%%%%%%%%%

%%%%%%%%%%%%%%%%%%%%%%%%%%%%%%%%%%%%%%%%%%%%%%%%%%%%%%%%%%%%%%%%%
\section{The calibration of the dredge-up parameters for the LMC}
\label{calibr}

The dredge-up parameters, $T_{\rm b}^{\rm dred}$ and $\lambda$, need
to be calibrated on the basis of some observational constraint. First,
we aim at reproducing the CSLF in the LMC, which has already been the
subject of previous works (e.g.\ Groenewegen \& de Jong 1993; Marigo
et al.\ 1996a). To this aim, we employ the results of synthetic
calculations for TP-AGB stars with initial chemical composition
$[Y=0.250; Z=0.008]$.  The adopted metallicity is in agreement with
the observational determinations for the LMC, specially with regard to
the young components (e.g.\ Russell \& Dopita 1990; Dopita et al.\
1997). Moreover, Olszewski et al.\ (1991) and de Freitas Pacheco et
al.\ (1998) find similar metallicities in the LMC star clusters
younger than about 4 Gyr.

By means of the method outlined in Sect.~\ref{theo_cslf}, we have
calculated the CSLF for different combinations of the two
parameters, while keeping fixed the other input prescriptions (namely
the IMF and SFR).  Specifically, we employed Salpeter's law (1955)
for the IMF, and the history of SFR suggested by Bertelli et al.\
(1992) in their study of the CMDs of field stars in selected areas of
the LMC. The SFR has been moderate from the beginning up to about 5
Gyr ago and since then a factor of ten stronger (see the top panel of
Fig.~\ref{sfr3}). The sensitivity of the results to various possible
choices of the SFR as a function of the age, and to the assumed IMF,
are explored in Sect.~\ref{sfr}.

In Fig.~\ref{cstar3} we show the predicted LF of carbon stars for
three pairs of dredge-up parameters.  The values are the following:
\begin{description}
\item $\log T_{\rm b}^{\rm dred} = 6.5$,  $\lambda=0.65$ 
\item $\log T_{\rm b}^{\rm dred} = 6.4$,  $\lambda=0.65$ 
\item $\log T_{\rm b}^{\rm dred} = 6.4$,  $\lambda=0.50$ 
\end{description}
The corresponding transition luminosities from the oxygen-rich to the
carbon-rich domain are plotted as a function of the initial mass in
Fig.~\ref{limbolz008}.  It is apparent that the formation of carbon stars
of lower masses is favoured at decreasing $T_{\rm b}^{\rm dred}$ and
increasing $\lambda$.  On the contrary, the luminosity at the tip of
the TP-AGB seems to be essentially independent of the dredge-up
parameters (at least for the cases under consideration).  A reliable
calibration of these parameters requires one to disentangle which distinctive
effects are produced by one parameter without the influence of the
other. This because it would be desirable to derive a unique pair of
proper values, or at least to reduce the suitable space of parameters
as much as possible.

\subsection{The temperature parameter: $T_{\rm b}^{\rm dred}$}
\label{tbmax}

%%%%%%%%%%%%%%%%%%%%% FIGURE  %%%%%%%%%%%%%%%%%%%%%%
%
\begin{figure}
\resizebox{\hsize}{!}{\includegraphics{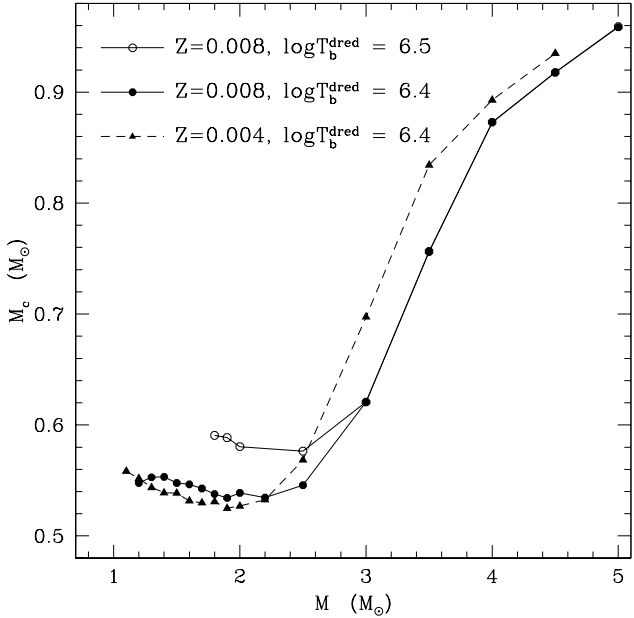}}
\caption{Core mass at the first dredge-up event as a function of the
stellar mass at the onset of the TP-AGB phase, for two values of the
temperature parameter $T_{\rm b}^{\rm dred}$, and metallicities 
$Z=0.008$ and $Z=0.004$.}
\label{mcdredz008}
\end{figure}
%
%%%%%%%%%%%%%%%%%%%%%%%%%%%%%%%%%%%%%%%%%%%%%%%%%%%%%%%%%%%

Let us first discuss the role of $T_{\rm b}^{\rm dred}$.  We recall
that this parameter determines the minimum core mass for dredge-up,
$M_{\rm c}^{\rm min}$, as a function of the stellar mass and
metallicity.

In a previous analysis (Marigo et al.\ 1996a) the reproduction of the
CSLF in the LMC was obtained adopting $M_{\rm c}^{\rm min}=0.58
M_{\odot}$ as a constant parameter, for all stellar masses and
chemical compositions (see also Groenewegen \& de Jong 1993 for
similar results).  In contrast, we have shown in Sect.~\ref{mcmin}
that with the new algorithm based on envelope integrations, at fixed
$T_{\rm b}^{\rm dred}$ the behaviour of $M_{\rm c}^{\rm min}$ as a
function of $M$ is far from being constant, but steeply increasing for
lower values of the mass.  Moreover, the onset of the third dredge-up
is favoured at lower $Z$.

Figure~\ref{mcdredz008} illustrates the core mass -- necessarily $\ge
M_{\rm c}^{\rm min}$ -- just before the occurrence of the first
dredge-up episode as a function of the total mass at the starting of
the TP-AGB phase for the evolutionary models calculated with initial
metallicity $Z=0.008$ and $Z=0.004$.  The comparison with the previous
Fig.~\ref{mcminz008} shows the following  differences.
	\begin{itemize}
	\item
First, the core mass at the first dredge-up event does not increase
so steeply as $M_{\rm c}^{\rm min}$
in the range of low-mass stars $M<2.5 M_\odot$.  This is the result of
the extreme sensitivity of $M_{\rm c}^{\rm min}$ to the current
stellar mass.  Even small reduction of the envelope actually
prevents the formation of carbon stars in the steepest part of the
curve shown in Fig.~\ref{mcminz008}.
	\item 
Second, the core mass at the onset of the third dredge-up is found to
increase for stars with $M\ga2.5 M_\odot$.  In this mass range, it
reflects the behaviour of the core mass at the first thermal pulse
(which increases with stellar mass), and not the flattening predicted
for $M_{\rm c}^{\rm min}$.
	\item
Finally, the core masses in the low-mass range 
indicated in Fig.~\ref{mcdredz008} are
somewhat larger than the corresponding ones shown in
Fig.~\ref{mcminz008}.  The reason is partly due to the fact 
that the results illustrated in
this latter plot are derived from calculations in which $L_{\rm P} =
L_{(\rm P,full)} + \Delta L_{(\rm P,first)}$ (see Sect.~\ref{lpeak}),
whereas the term $\Delta L_{(\rm P,first)}$ has been neglected in the
envelope integrations performed to construct
Fig.~\ref{mcminz008}. This simplification was adopted because $\Delta
L_{(\rm P,first)}$ contains a dependence on the core mass at the first
thermal pulse [see Eq.~(\ref{dlp})], whereas at that stage 
we simply aimed at
deriving the behaviour of $M_{\rm c}^{\rm min}$ as a function of the
current stellar mass.  The neglect of $\Delta L_{(\rm P,first)}$
leads one to overestimate $L_{\rm P}$ for a given core mass, thus yielding
slightly lower values of $M_{\rm c}^{\rm min}$.
	\end{itemize}

Comparing the curves for the same $\log T_{\rm b}^{\rm dred}=6.4$
and metallicities $Z=0.008$ and $Z=0.004$, it turns out that
at decreasing metallicity the core mass at the first dredge-up episode
is smaller in the low-mass range ($M \la 2.2 M_{\odot}$), whereas it is
larger at higher stellar masses (($M \ga 2.2 M_{\odot}$).
This is explained as following.
The fact that at lower metallicities the minimum
base temperature for dredge-up is attained at lower values of $L_{\rm
P}$ (and hence $M_{\rm c}$) is only relevant in the low-mass domain,
since the onset of the third dredge-up at larger stellar masses
essentially reproduces the trend of the core mass at the first thermal
pulse, which actually increases at decreasing metallicities for given
stellar mass.

%It is interesting to notice that the near constancy of the core mass
%at the onset of dredge-up, found in the present analysis,
%qualitatively agrees with the a priori criterion of a constant $M_{\rm
%c}^{\rm min}$, adopted in previous works (Groenewegen \& de Jong;
%Marigo et al.\ 1996a). 

\begin{table}
\caption{Minimum initial mass, $M_{\rm min}^{\rm carb}$, of carbon
stars' progenitors, and age of the relative stellar generation for
different values of the temperature parameter $T_{\rm b}^{\rm dred}$
and original metallicities. The corresponding mass at the onset of the
TP-AGB is also indicated inside parentheses.}
\label{mincarbz008}
\centering
\begin{tabular}{cccc}
\hline
\noalign{\smallskip}
$Z$ &
$\log T_{\rm b}^{\rm dred}$ &
$M_{\rm min}^{\rm carb}$ &
Age (yr) \\
\noalign{\smallskip}
\hline 
\noalign{\smallskip}
0.008 & 6.7 &  2.00 (2.0) & $1.211\times 10^{9}$ \\
      & 6.5 &  1.83 (1.8) & $1.514\times 10^{9}$ \\
      & 6.4 &  1.32 (1.2) & $3.890\times 10^{9}$ \\
0.004 & 6.4 &  1.13 (1.1) & $5.220\times 10^{9}$ \\
\noalign{\smallskip}
\hline 
\noalign{\smallskip}
\end{tabular}
\end{table}

Moreover, 
if we now compare the two curves plotted in
Fig.~\ref{mcdredz008}, for two values of $T_{\rm b}^{\rm dred}$ 
and and the same $Z=0.008$, it
turns out that the temperature parameter determines the minimum mass
for a TP-AGB star to experience chemical pollution by convective
dredge-up.  The lower is $T_{\rm b}^{\rm dred}$ the lower is this
limit mass.  Then, for a given mass-loss prescription applied to the
previous evolution (i.e.\ during the RGB and part of the AGB up to the
onset of dredge-up), we can derive the corresponding stellar mass at
the ZAMS.  In other words, $T_{\rm b}^{\rm dred}$ fixes a lower limit
to the minimum initial mass, $M_{\rm min}^{\rm carb}$, for a star to
become a carbon star, and hence an upper limit to the oldest age of
the stellar population possibly contributing to the observed
distribution (see Table \ref{mincarbz008}).  

Another interesting point to be stressed is that the results shown in
Fig.~\ref{mcdredz008} are determined by the temperature parameter only
(since $\lambda$ cannot have any effect prior to dredge-up).  
This fact allows us to predict already
that the value $\log T_{\rm b}^{\rm dred} = 6.5$ should be ruled out,
otherwise we would miss the low-luminosity tail of the LMC 
CSLF.  Let us consider the minimum (with $M_{\rm c} = 0.577
M_{\odot}$) of the corresponding curve in Fig.~\ref{mcdredz008},
referring to a star with $M = 2.5 M_{\odot}$.  Even in the most
favourable (but quite unlikely) hypothesis, that the transition to
the C-class occurs soon after the first dredge-up episodes
(this would require $\lambda \sim 1$), the corresponding luminosity
derived from the $M_{\rm c}-L$ relation would be
$\log (L/L_\odot) \sim 3.706$.  This is equivalent to $M_{\rm
bol} \sim -4.564$, and even adding $1.25$ mag (i.e.\ the typical depth
of the low-luminosity dip; see Sect.~\ref{dip}), we get a value that
is more than $\sim 0.3$ mag brighter than the observed faint end of
the distribution for the LMC, at $M_{\rm bol} \sim -3$.

In the case of $\log T_{\rm b}^{\rm dred} = 6.4$, the minimum of the
curve is at $M=1.8 M_{\odot}$, with $M_{\rm c} = 0.538 M_{\odot}$.
Applying the same reasoning, we would obtain the faintest carbon stars
at $M_{\rm bol} \sim -2.795$.  The latter value is consistent
with observed faint end of the CSLF in the LMC.

Even if a full discussion on the carbon star transition luminosity as
a function of the stellar mass cannot be developed without considering
the combined effect of $T_{\rm b}^{\rm dred}$ and $\lambda$, we can
conclude from the previous analysis that {\it the temperature
parameter, $T_{\rm b}^{\rm dred}$, mainly determines the faint end of
the CSLF}.

\subsection{The efficiency parameter: $\lambda$}
\label{lambdapar}

Let us now analyse the role of the dredge-up efficiency parameter
$\lambda$.  We remind that $\lambda=0.65$ was the suitable value
estimated by Marigo et al.\ (1996a) such that, together with $M_{\rm
c}^{\rm min} = 0.58 M_{\odot}$, the CSLF in the LMC was
reproduced.  The top and middle panels of Fig.~\ref{cstar3} show the
theoretical distributions of carbon stars adopting $\lambda = 0.65$,
for $\log T_{\rm b}^{\rm dred}=6.5$ and $\log T_{\rm b}^{\rm
dred}=6.4$, respectively. None of them fits the observed histogram.

In the case with ($\log T_{\rm b}^{\rm dred} = 6.5$, $\lambda=0.65$),
the theoretical distribution is systematically shifted to brighter
luminosities, with respect to the observed one. The opposite situation
is found with the combination ($\log T_{\rm b}^{\rm dred} = 6.4$,
$\lambda=0.65$), which produces the peak of the predicted CSLF at a
fainter magnitude than observed, with an excess of carbon stars at
lower luminosity, and conversely a deficit in the brighter domain.
Note, however, that contrary to the previous result with ($\log T_{\rm
b}^{\rm dred} = 6.5$, $\lambda=0.65$), both extremes of the observed
distribution are fitted.

This circumstance suggests that the we are only required to shift the
peak of the histogram into the right luminosity bin.  We find that
this can be obtained by decreasing the parameter $\lambda$ while
keeping fixed $\log T_{\rm b}^{\rm dred} = 6.4$, rather than varying
the temperature parameter itself.  There are two main reasons.  First,
the constraint set by the faint end of the distribution, which has
been already fulfilled with the choice $\log T_{\rm b}^{\rm dred} =
6.4$, would be violated with a higher value of $T_{\rm b}^{\rm dred}$.
Second, we expect that a suitable decrease of $\lambda$ could move the
peak into the right location, and reproduce the faint tail of the
distribution at the same time. In fact, the transition
luminosity for a star with a very low-mass envelope ($\sim 1
M_{\odot}$) is only slightly affected by changes of $\lambda$
(provided that $\lambda\ga0.4)$, since $2-3$ dredge-up episodes
generally suffice to cause the transition to carbon star.

Concluding this discussion, we remark that {\it once $T_{\rm b}^{\rm
dred}$ has been determined, the efficiency parameter $\lambda$
essentially controls the peak location of the CSLF.}  We find that
{\it the best fit to the observed CSLF in the LMC is obtained adopting
$\log T_{\rm b}^{\rm dred} = 6.4$ and $\lambda=0.5$}. This fit is
shown in the bottom panel of Fig.~\ref{cstar3}.

%%%%%%%%%%%%%%%%%%%%%%%%%%%%%%%%%%%%%%%%%%%%%%%%%%%%%%%%%%%%%%%%%
\subsection{Sensitivity to the SFR and IMF}
\label{sfr}

The relative weight of different stellar generations contributing to
the observed CSLF is determined by the SFR history  in the host
galaxy, and by the IMF.  In this section, we investigate the
sensitivity of the predictions on both functions.  Let us first
analyse the effect due to the SFR history, while fixing the IMF equal
to Salpeter's (1955) prescription.  
Then, we explore the sensitivity of the CSLF to different choices of
the IMF.

We now address the following questions:
\begin{itemize}
\item
To what extent is the CSLF sensitive to the underlying SFR ?
\item
Are there any distinctive features of the CSLF setting 
constraints on the history of SFR ?
\item 
What is more important in shaping the observed CSLF, the efficiency of the
dredge-up process or the time evolution of the SFR ?
\end{itemize}

%%%%%%%%%%%%%% FIGURE 5  %%%%%%%%%%%%%%%%%%%%%%%%%%%%%%%
%
\begin{figure}
\resizebox{\hsize}{!}{\includegraphics{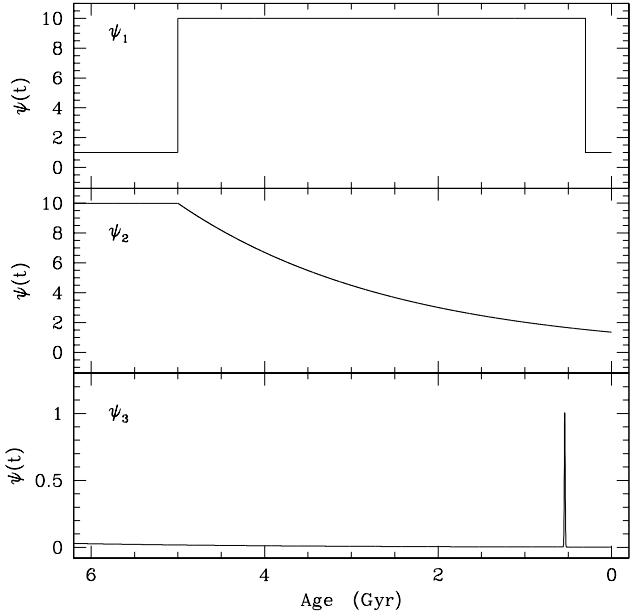}}
\caption{Schematic representation of three different prescriptions for
the SFR history in the LMC (in arbitrary units).  In the top panel the
law suggested by Bertelli et al.\ (1992) is shown [$\psi(t)=\psi_1$].
See the text for more details.}
\label{sfr3}
\end{figure}
%
%%%%%%%%%%%%%%%%%%%%%%%%%%%%%%%%%%%%%%%%%%%%%%%%%%%%%%%%%%

First of all, it is clear that the possible indications the CSLF may
give on the history of the SFR are restricted to the age interval
defined by the minimum and maximum  initial mass of the 
progenitors ($M_{\rm min}^{\rm carb}$ and $M_{\rm max}^{\rm carb}$, 
respectively). 
Specifically, we
deal with the age interval ranging from about $3.9 \times 10^{9}$ yr
down to $1.1 \times 10^{8}$ yr, corresponding to stellar lifetimes for
$M_{\rm min}^{\rm carb} = 1.32 M_{\odot}$ and $M_{\rm max}^{\rm carb}
= 5 M_{\odot}$, respectively. The lower limit of the mass range is
determined by the temperature parameter, $\log T_{\rm b}^{\rm
max}=6.4$ (see Sect.~\ref{tbmax}).

%%%%%%%%%%%%%% FIGURE 5  %%%%%%%%%%%%%%%%%%%%%%%%%%%%%%%
%
\begin{figure}
\resizebox{\hsize}{!}{\includegraphics{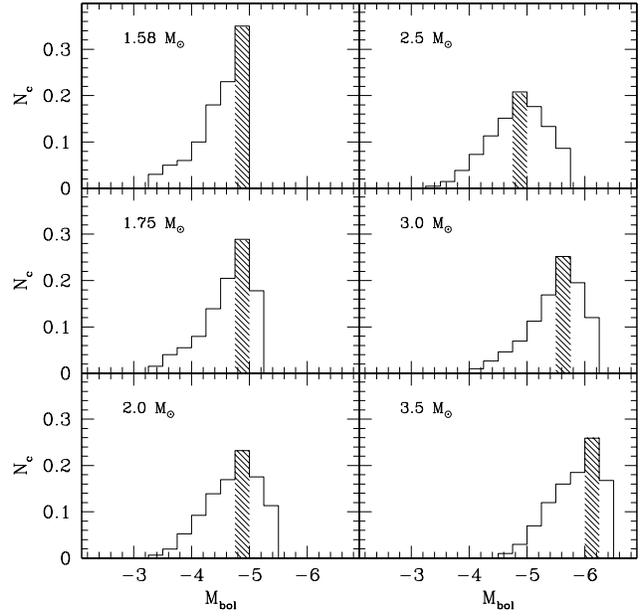}}
\caption{Theoretical CSLFs for simple stellar populations.  Each
histogram corresponds to the expected distribution of carbon stars of
the same age, i.e.\ evolved from progenitors with the same initial
mass, as indicated. The shaded areas mark the position of the
luminosity peak for each mass. Notice that all the histograms for
$\mi\le 2.5 M_{\odot}$ have the peak at the same luminosity
bin. Calculations refer to TP-AGB models of initial metallicity
$Z=0.008$, for values of the dredge-up parameters, $\log T_{\rm
b}^{\rm dred} = 6.4$ and $\lambda=0.5$. Similar behaviours hold for
other choices of the parameters.}
\label{cstar_mass}
\end{figure}
%
%%%%%%%%%%%%%%%%%%%%%%%%%%%%%%%%%%%%%%%%%%%%%%%%%%%%%%%%%%%

An interesting result comes out when comparing the theoretical
distributions, each corresponding to a single stellar generation, as a
function of the initial mass of the progenitor, or equivalently, of
the age. These theoretical CSLFs are shown in Fig.~\ref{cstar_mass}, for
the case $\log T_{\rm b}^{\rm dred} = 6.4$ and $\lambda=0.5$. For
each stellar mass (and hence age), the CSLF has a shape
determined essentially by the convolution of a box-like function (see
Sect.~\ref{ratebri}) with the probability function described in
Sect.~\ref{dip}.  From this figure, it turns out that all the
histograms relative to stars with initial masses $\mi\le 2.5 M_{\odot}$
have a faint wing extending within the same luminosity range, and more
importantly, all the peaks fall into the same bin.  This is explained
considering that, within this mass range, (i) stars are found to make
the transition to the C-class at quite similar luminosities (i.e.\
$M_{\rm bol} \sim -4$; see Fig.~\ref{limbolz008}), and (ii) they span
as carbon stars a luminosity interval which is comparable to the
typical excursion of the low-luminosity dip, i.e.\ $\Delta M_{\rm bol}
\sim 1$ mag.

In view of this similarity, we expect that the faint wing of the
integrated CSLF is not significantly affected by possible variations in
the SFR occurring at ages older than $\sim 7 \times 10^{8}$~yr and, in
particular, the location of the peak would remain unchanged.  On the
other hand, differences are evident among the distributions relative
to more massive ($\mi> 2.5 M_{\odot}$) carbon stars. Hence, it follows
that the high luminosity tail could reflect the details of the recent
history of star formation, for ages comprised roughly within $7\times
10^{8}$ and $10^{8}$ yr.

On the basis of the above considerations, let us now investigate
whether it is possible to reproduce the observed CSLF by
only varying the adopted SFR, while keeping fixed all the other
prescriptions.  We start from the theoretical distributions which
are found not to fit the observed constraints with our standard
prescription for the SFR [$\psi(t)=\psi_1$, top panel of
Fig.~\ref{sfr3}], for two different combinations of the dredge-up
parameters (top and middle panels of Fig.~\ref{cstar3}; see also
Sects.~\ref{tbmax} and \ref{lambdapar}).

%%%%%%%%%%%%% 1 case %%%%%%%%%%%%%%%%%%%%%%%%%%%%%%%%%%%%%%%%

Concerning the $(\log T_{\rm b}^{\rm dred} = 6.5, \lambda =0.65)$
case, shown in the top panel of Fig.~\ref{cstar3}, we can already
conclude that the discrepancy cannot be removed, i.e.\ there is no way
to shift the theoretical peak towards lower luminosities.  In fact, as
previously discussed, the faintest possible location of the peak is
characteristic of all simple distributions of carbon stars with
initial masses $\le 2.5 M_{\odot}$, corresponding to ages older than
$\sim 7 \times 10^{8}$~yr.  This circumstance makes the lower limit of
the peak luminosity almost invariant to any changes in the SFR
over this age interval.  We can then conclude that {\it the luminosity
location of the peak is indeed a strong calibrator of the efficiency
parameter $\lambda$}.

Moreover, considering that the extension of the faint tail of the
distribution, from the peak down to the end, is essentially controlled
by the occurrence of the low-lumino\-sity dip (see Sect.~\ref{dip}), it
follows that it not possible to populate the observed faintest
bins, except by invoking a larger extension of the subluminous stage
(e.g.\ $\sim 1.75$~mag) than assumed here ($1.25$~mag).

%%%%%%%%%%%%%% 2 case %%%%%%%%%%%%%%%%%%%%%%%%%%%%%%%%%%%%%%%%%%%%%%%%%%%%%%%%%

Let us now analyse the ($\log T_{\rm b}^{\rm dred} = 6.4$,
$\lambda=0.65$) case, illustrated in the middle panel of
Fig.~\ref{cstar3}. We aim at testing whether it is possible to
calibrate a suitable law for the SFR, which is able to make the
theoretical peak coincide with the observed one and to attain a
general agreement of the overall features.  After several trials, the
best fit is obtained by approximating the SFR as the sum of a
decreasing exponential and a gaussian function:
\begin{equation}
\psi(t) \propto \exp\left(-\frac{t}{t_{\rm s}}\right)+ 
\exp\left[-\frac{1}{2} \left(
	\frac{t-\overline{t}}{\sigma_{t}}\right)^{2}\right] 
\end{equation} 
which is illustrated in the bottom panel of Fig.~\ref{sfr3} [$\psi(t)
= \psi_{3}$].  The suitable value for the time-scale parameter is
$t_{\rm s} = 2.5$~Gyr, producing quite a slow decline of the SFR towards
younger ages.  The effect of the slight decrease is not negligible, as
it secures the reproduction of the faint end of the distribution, not
reached otherwise with a strictly constant SFR.

%%%%%%%%%%%%%% FIGURE 6  %%%%%%%%%%%%%%%%%%%%%%%%%%%%%%%
%
\begin{figure}
\resizebox{\hsize}{!}{\includegraphics{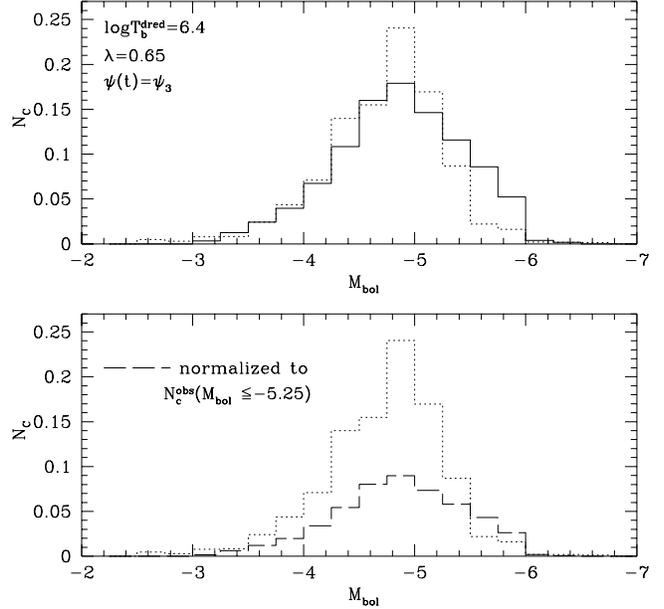}}
\caption{Top panel: the theoretical LF of carbon stars (solid line)
obtained with the dredge-up parameters as indicated, and a
prescription for the SFR [$\psi(t)=\psi_{3}$] characterised by a
recent burst at around $5.5 \times 10^{8}$~yr ago.  The dotted line
histogram represents the observed LF of carbon stars in the LMC.
Bottom panel: The same as in the top panel, but with the theoretical
histogram (dashed line) normalised to the number of observed carbon
stars more luminous than $M_{\rm bol} = -5.25$.}
\label{sfr_t64l65}
\end{figure}

%%%%%%%%%%%%%%%%%%%%%%%%%%%%%%%%%%%%%%%%%%%%%%%%%%%%%%%%%%%

The gaussian function, centred at the age $\overline{t} = 5.5 \times
10^{8}$~yr and with a standard deviation $\sigma_{t} = 0.005$, mimics
the occurrence of a strong and short-lived burst around few $10^{8}$
yr ago.  This favours the contribution from carbon
stars belonging to relatively young stellar generations ($\mi \sim 2.8
M_{\odot}$), so that the peak of the theoretical distribution
is moved into the right luminosity bin.  However, it is worth noticing
that, although the overall luminosity range fits the observed one, an
excess of luminous carbon stars ($M_{\rm bol} <- 5.25 - -6$) is
predicted.  To get a quantitative estimate of corresponding error we
also plot in Fig.~\ref{sfr_t64l65} the same theoretical distribution, but
normalized to the total number of observed carbon stars brighter than
$M_{\rm bol } = -5.25$.  In this case, the absolute agreement at high
luminosities reflects into a considerable deficit of carbon stars
fainter than $M_{\rm bol } > -5.25$, covering a fraction of $\sim 37
\%$, which is considerably smaller than the observed fraction, $\sim
87 \%$.  The mismatch is really significant.

%%%%%%%%%%%%% 3  case %%%%%%%%%%%%%%%%%%%%%%%%%%%%%%%%%%%%%%%%%%%%%%%%%%%%

Finally, let us consider the $(\log T_{\rm b}^{\rm dred} = 6.4$,
$\lambda =0.5)$ case, shown in the bottom panel of Fig.~\ref{cstar3},
which is the best fit to the observed CSLF in the LMC.  The
theoretical distribution is obtained adopting a step-wise SFR
[$\psi(t)=\psi_1$ in Fig.~\ref{sfr3}], that corresponds to the
occurrence of a dominant episode of star formation since the age
$t_{\rm b}=5 \times 10^9$~yr, up to some more recent epoch $t_{\rm
e}$.

In practice, within the significant age range for the formation of
carbon stars, $\psi(t)$ is constant up to $t_{\rm e}$, when the SFR is
assumed to suddenly drop to very low (but not zero) values. From our
analysis, it turns out that the best agreement with the observed
high luminosity tail of the distribution ($M_{\rm bol} < -5.6$) is
obtained with $t_{\rm e} \sim 5 \times 10^{8}$~yr.  Since the age
defining the end of the burst is higher than that of the most massive
(i.e.\ $4 - 5 M_{\odot}$) carbon stars, their contribution remains
quite modest, but sufficient to account for the bright end of the
distribution at $M_{\rm bol} \sim -6.8$.

If younger ages are assigned to $t_{\rm e}$, a slight excess of
luminous carbon stars would show up. In this case, the discrepancy may
be removed invoking, instead of a recent sharp decrease in the SFR, a
more efficient envelope burning, or stronger mass-loss rates in
massive TP-AGB stars. In this respect, it is worth noticing the high
luminosity bins of the distribution are crucially affected by the
delicate interplay between convective dredge-up, envelope burning, and
mass-loss occurring in massive TP-AGB stars. Unfortunately, the
present understanding of all these topics is incomplete so that
theoretical predictions suffer from a  degree of uncertainty.

\begin{figure}
\resizebox{\hsize}{!}{\includegraphics{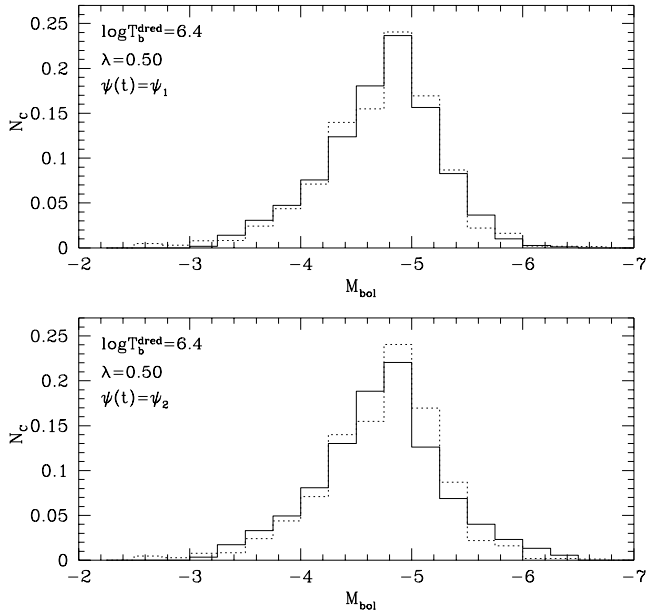}}
\caption{Top panel: the best fit (solid line) to the observed CSLF
(dotted line) in the LMC.  The theoretical distribution is obtained
with the adoption of the dredge-up parameters as indicated, and
following the prescription for the SFR ($\psi(t)=\psi_{1}$) suggested
by Bertelli et al.\ (1992).  Bottom panel: The same as in the top
panel, but assuming a SFR slowly decreasing with time
[$\psi(t)=\psi_{2}$; see also Fig.~\protect\ref{sfr3}].}
\label{sfr_t64l50}
\end{figure}

A further test is performed adopting a SFR law that is slowly
decreasing -- rather than constant -- from an age of $5$~Gyr up to
now [$\psi(t)=\psi_2$, middle panel of Fig.~\ref{sfr3}]:
\begin{equation}
\psi(t) \propto \exp(-t/t_{\rm s})
\end{equation}
with a time-scale $t_{\rm s} = 2.5$~Gyr.  The results, shown in
Fig.~\ref{sfr_t64l50}, are satisfactory for the faint wing of the
distribution, but both the peak and the high luminosity tail are
somewhat less well reproduced than with the $\psi(t) = \psi_1$
prescription.

%%%%%%%%%%%%%% FIGURE 7  %%%%%%%%%%%%%%%%%%%%%%%%%%%%%%%
%
%
%%%%%%%%%%%%%%%%%%%%%%%%%%%%%%%%%%%%%%%%%%%%%%%%%%%%%%%%%%%

From the above investigation it follows that {\it the observed faint
tail and the peak location of the CSLF cannot give stringent limits on
the history of the SFR, as they are essentially determined by the
dredge-up parameters.  On the other hand, the bright wing of the
distribution is sensitive to the details of the recent SFR}.

%%%%%%%%%%%%%% FIGURE 8 %%%%%%%%%%%%%%%%%%%%%%%%%%%%%%%
%
\begin{figure}
\resizebox{\hsize}{!}{\includegraphics{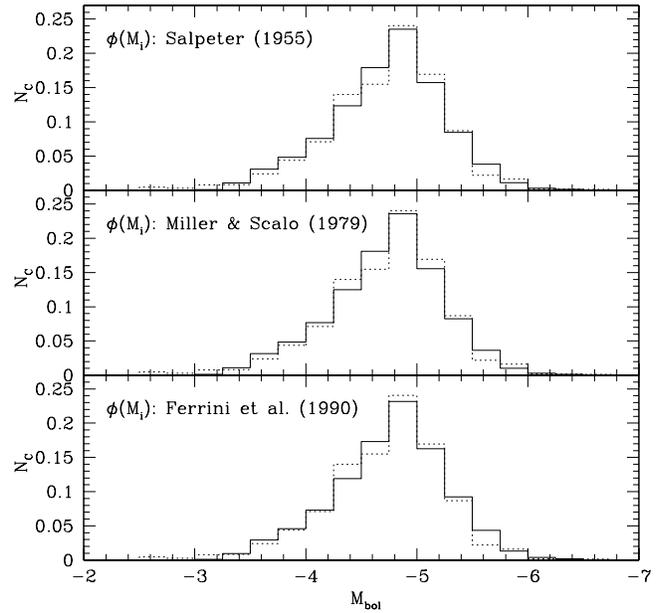}}
\caption{Predicted CSLFs according to different prescriptions for the
IMF, as quoted in each panel (solid lines).  The dotted line shows the
observed distribution in the LMC.  The adopted dredge-up parameters
are $(\log T_{\rm b}^{\rm dred} = 6.4$, $\lambda =0.5)$, and the SFR
is $\psi_{1}$.
}
\label{imf}
\end{figure}
%
%%%%%%%%%%%%%%%%%%%%%%%%%%%%%%%%%%%%%%%%%%%%%%%%%%%%%%%%%%%

Let us now explore the sensitivity of the result to the IMF.  In
Fig.~\ref{imf} we present the resulting distributions of carbon stars
adopting our best fit for the SFR (i.e. $\psi_{1}$ shown in the
top-panel of Fig.~\ref{sfr3}), while varying the law for the IMF
according to three choices, namely: Salpeter (1955); Miller \& Scalo
(1979); and Ferrini et al.\ (1990).  The first two prescriptions for
the IMF are empirical, while the third one has a theoretical
derivation.

The differences among the predicted luminosity functions are indeed
negligible.  This can be understood considering that:
	\begin{itemize}
	\item the similarity of the predicted distributions for simple
stellar populations with turn-off masses $1 \la (M_{\rm i}/M_{\odot})
\la 2.5$ makes the results quite insensitive to the adopted IMF in
this mass range, which indeed provides the major contribution of
carbon stars;
	\item the small weight of the IMF at higher masses tends to
cancel the differences in the results obtained with different
prescriptions.
	\end{itemize}

In conclusion, {\it it turns out that the observed distribution
of the carbon stars in the LMC is mainly determined by the 
the properties of the third dredge-up (mass-loss, and envelope
burning), with the SFR and IMF producing a much weaker effect.}

%%%%%%%%%%%%%%%%%%%%%%%%%%%%%%%%%%%%%%%%%%%%%%%%%%%%%%%%%%%
\section{The calibration of dredge-up parameters for the SMC}
\label{smc}

%%%%%%%%%%%%%%%%%%%% FIGURE %%%%%%%%%%%%%%%%%%%%%%%%%%%%%%%%%%%%%%
\begin{figure}
\resizebox{\hsize}{!}{\includegraphics{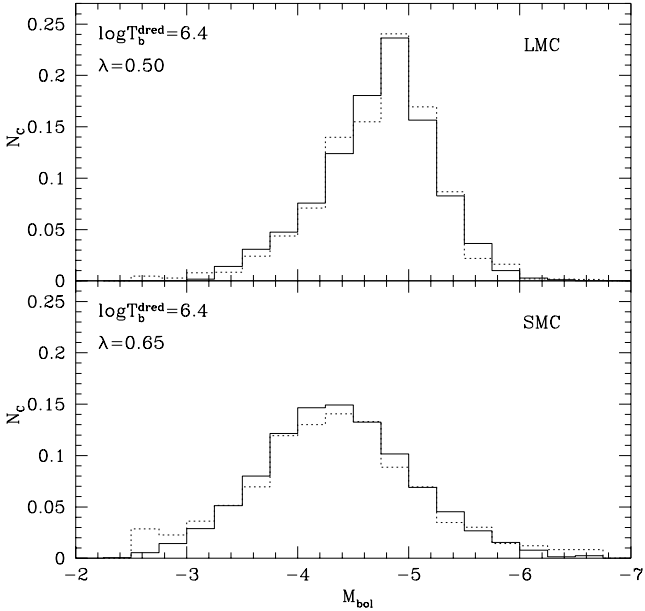}}
\caption{Observed CSLFs (dotted line) in the LMC (top
panel) and SMC (bottom panel), and the theoretical best fits (solid
line).}
\label{slmc}
\end{figure}
%%%%%%%%%%%%%%%%%%%%%%%%%%%%%%%%%%%%%%%%%%%%%%%%%%%%%%%%%%%%%%%%%%

The same kind of procedure described in the previous section is also
applied to reproduce the CSLF in the SMC.  In this case, we made use
of synthetic calculations for TP-AGB stars with initial chemical
composition $[Y=0.240; Z=0.004]$. This values are consistent with
present data for the abundances of young populations in the
SMC. However, it is also found that the SMC clusters present a clear
age--metallicity relation (Da Costa \& Hatdizimitriou 1998), with
older clusters being systematically more metal poor. If we adopt the
mean age--metallicity relation defined by the latter authors (i.e.\
excluding the apparently abnormal clusters L~113 and NGC~339), we find
that it is necessary to go to ages as old as 5~Gyr (corresponding to C
stars with $\mi\sim1.25\;M_\odot$) to find abundances reduced by a
factor of 2. Therefore, it is possible that $Z=0.002$ evolutionary
tracks would be more suitable to represent the oldest (and faintest)
SMC carbon stars. We remark, however, that the present investigation
is intended to give a first hint about the behaviour of the dredge-up
parameters at metallicities lower than $Z=0.008$. Relevant transition
luminosities as a function of the initial mass are displayed in
Fig.~\ref{limbolz004}, for two choices of the dredge-up parameters.

The best fit is obtained adopting $(\log T_{\rm b}^{\rm dred} = 6.4,\;
\lambda =0.65)$ in the simple case of a constant SFR and Salpeter's
IMF as shown in the bottom panel of Fig.\ref{slmc}.  It is interesting
to notice that the proper value of $T_{\rm b}^{\rm dred}$ is the same
as for the LMC. This is in agreement with the indications from
complete evolutionary calculations already discussed in
Sect.~\ref{tdred}, that $T_{\rm b}^{\rm dred}$ is constant regardless
of core mass and metallicity. 

Another aspect worthy of notice is that
a greater efficiency of the third dredge-up is required, if compared
to the results for the LMC.  This also agrees qualitatively with the
indications from complete modelling of thermal pulses (e.g.\ Boothroyd
\& Sackmann 1988d) that a deeper envelope penetration is favoured at
lower metallicities.
As already mentioned, the effect of increasing $\lambda$ is that of
shifting the peak location towards fainter luminosities.  Indeed, this
is one of the most evident differences between the distributions of
carbon stars in the two Clouds.

Concerning the broader extension in luminosity of the CSLF in the SMC if
compared to the LMC, the following effects are remarked:
\begin{itemize}
\item 
The faint end of the distribution is reached at lower luminosities in
the case of lower metallicities mainly because of both an earlier
onset of the third dredge-up, and the smaller number of dredge-up
episodes required for the transition to the C-class (due to the lower
original abundance of oxygen in the envelope).
\item 
As far as the bright end of the distribution is concerned, it is worth
noticing that a good fit to the observed data for the SMC is obtained
assuming a constant SFR over the entire significant age interval,
without the recent drop (at an age of $\sim 5 \times 10^{8}$ yr)
invoked for the LMC.
\end{itemize}

The CSLFs for both Magellanic Clouds are presented in the two panels
of Fig.~\ref{slmc}. It is remarkable that the present formulation for
the third dredge-up allows a good fitting of the CSLFs of both Clouds
without any dramatic change in the input parameters. In this context,
{\em the large difference between the CSLFs in the LMC and SMC are
interpreted mainly as the result of a different metallicity}. The
different histories of SFR in both galaxies probably play only a minor
role in their CSLFs.

%%%%%%%%%%%%%%%%%%%%%%%%%%%%%%%%%%%%%%%%%%%%%%%%%%%%%%%%%%%%%%%%%
\section{Conclusions}
\label{conclu}

In this paper we describe an improved treatment of dredge-up in
synthetic TP-AGB models, based on the adoption of the parameter
$T_{\rm b}^{\rm dred}$, i.e.\ the minimum base temperature required
for dredge-up to occur.  Every time a thermal pulse is expected during
calculations, envelope integrations are performed to check whether the
condition on the base temperature is satisfied. It follows that,
contrary to the test based on the constant $M_{\rm c}^{\rm min}$
parameter, with this new scheme the response depends not only on the
core mass but also on the current physical conditions of the envelope
(i.e.\ the surface luminosity peak, the effective temperature, the
mass, the chemical composition).  Moreover, this method allows one to
determine not only the onset, but also the possible shut-down of
dredge-up occurring when the envelope mass is significantly reduced by
stellar winds, without invoking further a priori assumptions.

This aspect is crucial.  In fact, after the shut-down of the third
dredge-up, surface and wind abundances of low-mass AGB stars are
expected to freeze out till the end of the evolution, with important
consequences for the chemical composition of planetary nebulae, which
are ejected during the last evolutionary stages.  Another point
concerns the competition between envelope burning and dredge-up in the
most massive TP-AGB stars. In particular, if the shut-down of
dredge-up occurs after envelope burning extinguishes, a star might
undergo a late conversion to carbon star.  The recent detection of a
luminous ($M_{\rm bol} \sim -6.8$) dust-enshrouded carbon star in the
LMC (van Loon et al.\ 1998) seems to indicate such eventuality. This
topic is the subject of a future study.

The CSLFs in the Magellanic Clouds are investigated as
possible calibrators of the dredge-up parameters, and possible
indicators of the SFR history in the host galaxies.  We briefly
summarise here the basic points emerging from our analysis:

	\begin{itemize}
	\item
In principle, the observed CSLFs can provide indications
on the SFR history within the age interval from about $4 \times
10^{9}$ yr to $\sim 10^{8}$ yr. Nothing can be inferred for younger
ages.
	\item 
The faint end of the luminosity distribution is essentially determined
by the temperature parameter $T_{\rm b}^{\rm dred}$.
	\item 
The peak location depends only weakly on the SFR, and hence it is a
stringent calibrator of the efficiency parameter $\lambda$.
	\item 
Although the peak location is essentially invariant with respect to the SFR,
the shape of the carbon star distribution may give hints on the
history of the SFR in recent epochs (for ages between $7 \times
10^{8}$ and $10^{8}$~yr ago).  However, the bright wing is
expected to be sensitive to the details of
 physical processes occurring in the most
massive stars, i.e.\ envelope burning and mass loss.
	\item
The best fit to the observed CSLF in the LMC is obtained with
$\lambda=0.50$, $T_{\rm b}^{\rm dred} = 6.4$, and a constant SFR up to
an age of about $5 \times 10^{8}$ yr.  This recent drop of the SFR is
invoked to remove a slight excess of bright carbon stars otherwise
predicted.
	\item  
The best fit to the observed CSLF in the SMC is derived with
$\lambda=0.65$, $T_{\rm b}^{\rm dred} = 6.4$, and a constant SFR over
the entire significant age interval.
	\item 
If compared with the calibration for the LMC, a greater efficiency of
dredge-up is required to reproduce the fainter peak of the CSLF in the
SMC.
	\end{itemize}

%%%%%%%%%%%%%%%%%%%%%%%%%%%%%%%%%%%%%%%%%%%%%%%%%%%%%%%%%%%%%%%%%%%%%%%%%%%%%%

\begin{acknowledgements}
We are grateful to Cesare Chiosi for his 
important advice and kind interest in this work. We thank Achim Weiss
for carefully reading the manuscript and for his useful comments. Peter
Wood is acknowledged for the important suggestions which gave origin
to the treatment of dredge-up here proposed. 
Many thanks to our referee, Arnold I.\ Booth\-royd, 
whose remarks helped us to significantly 
improve the final version of this paper.
The work by L.\ Girardi
is funded by the Alexander von Humboldt-Stiftung.
A.\ Bressan acknowledges the support by the European Community under 
TMR grant ERBFMRX-CT96-0086.
\end{acknowledgements}

\end{document}